\documentclass[prc,twocolumn,superscriptaddress,showpacs,floatfix]{revtex4}
\usepackage{graphicx,amsmath,amssymb,bm}

\def\be{\begin{equation}}
\def\ee{\end{equation}}
\def\ba{\begin{eqnarray}}
\def\ea{\end{eqnarray}}
\def\bas{\begin{eqnarray*}}
\def\eas{\end{eqnarray*}}

\newcommand{\vlowk}{V_{{\rm low}\,k}}

\begin{document}

\title{{\it Ab initio} coupled-cluster approach to nuclear structure
  with modern nucleon-nucleon interactions}

\author{G.~Hagen}
\affiliation{Physics Division, Oak Ridge National Laboratory,
Oak Ridge, TN 37831, USA}
\author{T.~Papenbrock}
\affiliation{Department of Physics and Astronomy, University of  
Tennessee, Knoxville, TN 37996, USA}
\affiliation{Physics Division, Oak Ridge National Laboratory,
Oak Ridge, TN 37831, USA}
\author{D.~J.~Dean}
\affiliation{Physics Division, Oak Ridge National Laboratory,
Oak Ridge, TN 37831, USA}
\author{M.~Hjorth-Jensen}
\affiliation{Department of Physics and Center of Mathematics for  
Applications, University of Oslo, N-0316 Oslo, Norway}

\begin{abstract}
  We perform coupled-cluster calculations for the doubly magic nuclei
  $^4$He, $^{16}$O, $^{40,48}$Ca, for neutron-rich isotopes of oxygen
  and fluorine, and employ ``bare'' and secondary renormalized
  nucleon-nucleon interactions.  For the nucleon-nucleon interaction
  from chiral effective field theory at order next-to-next-to-next-to
  leading order, we find that the coupled-cluster approximation
  including triples corrections binds nuclei within 0.4 MeV per
  nucleon compared to data. We employ interactions from a
  resolution-scale dependent similarity renormalization group
  transformations and assess the validity of power counting estimates
  in medium-mass nuclei. We find that the missing contributions due to
  three-nucleon forces are consistent with these estimates. For the
  unitary correlator model potential, we find a slow convergence with
  respect to increasing the size of the model space.  For the
  $G$-matrix approach, we find a weak dependence of ground-state
  energies on the starting energy combined with a rather slow
  convergence with respect to increasing model spaces. We also analyze
  the center-of-mass problem and present a practical and efficient
  solution.
\end{abstract}


\maketitle

\section{Introduction}

In the last decade, ab-initio nuclear structure calculations have made
great
progress~\cite{Piep01,Nav00,Nogga,Pisa,DeanLee,Nav07,NCSM_Review,Hag08,Bacca09}.
Light nuclei up to carbon or so can now be described in terms of their
nucleonic degrees of freedom, realistic nucleon-nucleon (NN) forces
(i.e., those that fit the available body of NN data with a
$\chi^2\approx 1$ per datum) augmented by a three-nucleon force (3NF).
One of the major advances is due to the systematic construction of
nuclear forces within chiral effective field theory
(EFT)~\cite{Weinberg,Kolck94,Epel00,N3LO,N3LOEGM,machleidt02}. Within this
approach, the unknown short-range part of the nuclear force is
systematically encoded in terms of low-energy constants and contact
terms, while the long-range part of the interaction stems from pion
exchange. One of the hallmarks of this approach is the ``power
counting'', i.e., an expansion of the nuclear Lagrangian in terms of
the momentum ratio $Q/\Lambda_\chi$. Here, $Q$ denotes the typical momentum
scale at which the nucleus is probed, while $\Lambda_\chi$ denotes the
high-momentum cutoff scale that limits the applicability of the
effective field theory. Within this approach, three-nucleon forces
appear naturally at next-to-next-to-leading (N$^2$LO) order
$(Q/\Lambda_\chi)^3$, and four-nucleon forces get introduced at N$^3$LO,
that is at order $(Q/\Lambda_\chi)^4$ in terms of the momentum scale and
the cutoff $\Lambda_\chi$.

The chiral interactions have been tested in very light systems of mass
$A$=$3,4$ through precise few-body calculations \cite{Nogga,Pisa}, and
in $p$-shell nuclei within no-core shell model (NCSM)
calculations~\cite{Nav07,NCSM_Review}. In these calculations, the NN
interaction was taken up to N$^3$LO, while the three-nucleon
interaction was taken at its leading order $(Q/\Lambda_\chi)^3$ (N$^2$LO).
Within the NCSM calculations, the ``bare'' chiral interaction needs to
be renormalized, due to the size of the model space. Lattice
calculations provide a different approach for the implementation and
solution of chiral interactions in the nuclear many-body
problem~\cite{DeanLee}.  Presently, such calculations employ two- and
three-nucleon forces at order N$^2$LO, and they target $p$-shell
nuclei.  The coupled-cluster method is an alternative approach that is
particularly suited to study the saturation properties of chiral
nucleon-nucleon interactions in medium-mass nuclei such as
$^{40,48}$Ca and $^{48}$Ni~\cite{Hag08}, or even heavier nuclei.
Using a Gamow basis expansion~\cite{michel10}, this method can also
include continuum effects and describe rare isotopes that are in the
vicinity of nuclei with closed (sub)shells~\cite{HagenPLB,hagen10}.

The ``bare'' chiral interaction models are still relatively ``hard''
interactions, that is the interaction has nonzero matrix elements even
for high momentum transfers $Q \gg k_F$, where $k_F\approx 1.35$
fm$^{-1}$ denotes the Fermi momentum at nuclear saturation. Using for
example matrix elements of the chiral interaction by Machleidt and
Entem~\cite{N3LO}, we extend typically the momentum integrations up to
8~fm$^{-1}$. For a model space consisting of oscillator wave
functions, a simple estimate shows that the number of oscillator
shells required for the solution of a nucleus with radius $R$ and an
interaction with momentum cutoff $\lambda$ is about $N\approx \lambda
R$ (see Sect.~\ref{sec:medmass} below for details). Recall that the
number of single-particle states (single-particle $j$-shells) grows as
$N^3$ in the $m$-scheme (and as $N^2$ in a $j$-coupled spherical
scheme).  It is therefore clear that high-momentum interactions
require large model spaces. This makes wave-function-based solutions
of the nuclear many-body problem a challenging task. Within the
coupled-cluster approach, one can overcome this difficulty through a
spherically symmetric reformulation of the method. This approach
enables us to consider much increased model spaces, including even up
to 20 oscillator shells. The obvious alternative consists of lowering
the momentum cutoff.

The momentum cutoff of the interactions can be decreased by a
similarity renormalization group (SRG) transformation of the NN
interaction~\cite{SRG}, or by integrating out high-momentum modes
within a renormalization group transformation or a similarity
transformation~\cite{Vlowk1,Vlowk2}.  Such transformations preserve
all properties (such as phase shifts and bound states) of the NN
interactions up to the cutoff. The SRG transformations yield
interactions that are band diagonal in momentum space, and the band
width is characterized by a momentum scale.  Eigenstates of these
interactions are localized in momentum space, and the low-momentum
states are thus decoupled from the high-momentum physics.  The
low-momentum interactions $\vlowk$ are characterized by a
momentum cutoff beyond which all interaction matrix elements quickly
approach zero. We denote the momentum scale of the SRG interactions
and the cutoff of the $\vlowk$ interactions by $\lambda$.
Starting with $\lambda=\Lambda_\chi$, one can thus generate a one-parameter
family of interactions from the chiral NN interaction models by
performing SRG transformations, or by integrating out high-momentum
modes. We will use this approach in order to study the scale- or cutoff
dependence of the nuclear binding energy, and to examine aspects of
the power counting. Throughout this work we make the underlying
assumption that a complete description within chiral effective field
theory would yield an accurate description of atomic nuclei. Thus, we
attribute any scale dependence directly to the neglected three-nucleon
forces and other neglected high-order terms of the interaction.

The more traditional way of dealing with ``hard'' interactions
consists of the computation of the so-called $G$-matrix~\cite{mhj95},
which is based on a Green's function approach \cite{greens} using
normally unperturbed propagators. The $G$-matrix depends thus on the
starting energy that is employed in its construction. Analytical
arguments demand that this dependence weakens as increasingly larger
model spaces are considered, and it vanishes for infinite spaces.
Within the spherical coupled-cluster approach, we are able to
investigate the convergence properties of the $G$-matrix that is
constructed from chiral interaction models.

Another approach to renormalize ``hard'' interactions is the unitary
correlator method (UCOM). Here a unitary transformation is constructed
to remove the hard core and short-range contributions of the tensor
force by an appropriately formulated correlation operator~\cite{UCOM}.
This method has seen several applications, but its convergence and
saturation properties have not yet been studied in a framework that
includes substantial wave function correlations. We fill this gap in
this work.

It is the purpose of this paper to employ ``bare'' and renormalized
nucleon-nucleon interactions (employing SRG and low-momentum
techniques, the $G$-matrix and the UCOM method), to compare and
analyze their convergence and saturation properties and their impact
on nuclear structure. In addition to this task, this paper also
contains significant supplemental information that could not be
presented in several recent short
communications~\cite{Hag08,Dean08_Comment,Hag09_ox,Hagen_PRL2009,hagen10}.
In particular, we present a detailed study and practical solution of
the center-of-mass problem, and study the evolution of single-particle
energies in neutron-rich oxygen and fluorine isotopes.

This paper is organized as follows. In Section~\ref{sec:theory}, we
introduce spherical coupled-cluster theory. We address and resolve
questions regarding the center-of-mass problem in
Section~\ref{sec:com}. In Section~\ref{sec:medmass}, we compute the
binding energies of the nuclei $^4$He, $^{16}$O, $^{40}$Ca, and
$^{48}$Ca starting from ``bare'' chiral NN interactions.
Section~\ref{sec:evolve} is dedicated to the evolution of single
particle energies in neutron-rich oxygen isotopes. In
Section~\ref{sec:srg}, we study $^{40}$Ca for several momentum scales
$\lambda$ of the SRG interaction, and examine the power counting.
Section~\ref{sec:g} focuses on the starting-energy dependence and the
convergence properties of the $G$-matrix for $^{4}$He and $^{16}$O. We
analyze the convergence properties of the UCOM interaction in
Section~\ref{sec:ucom}. We finish with our Summary.  Some technical
details of the spherical coupled-cluster method are relegated to the
Appendix.

\section{Spherical coupled-cluster theory}
\label{sec:theory}
In this section we give an outline of the coupled-cluster method 
\cite{Coe58,Coe60,Ciz66,Ciz69,Kuem78,Bar07,Guardiola,Hei99,Mi00,Dean04,Kow04,Wlo05} 
and introduce an angular momentum coupled formulation of the
coupled-cluster equations. First we outline coupled-cluster
theory for the computation of ground state energy of closed-shell 
nuclei within the so-called CCSD and $\Lambda$-CCSD(T) approaches. Thereafter, 
we introduce 
the equation-of-motion theory for the calculation of ground and excited states
in closed- and open-shell nuclei. We introduce also the spherical 
formulation of coupled-cluster theory. Finally, we discuss how 
to calculate expectation values of observables in coupled-cluster theory. 

\subsection{Coupled-cluster theory for closed-shell nuclei}
Coupled-cluster theory is based on the similarity transformation 
\begin{equation}
\label{hbar}
\overline{H}=e^{- T} H e^T
\end{equation}
of the normal-ordered Hamiltonian $H$. Here, the Hamiltonian is
normal-ordered with respect to a product state $|\phi_0\rangle$ which
serves as a reference. Likewise, the
particle-hole cluster operator
\begin{equation}
\label{T}
T = T_1 + T_2 + \ldots + T_A
\end{equation}
is defined with respect to the reference state. The $k$-particle 
$k$-hole ($k$p-$k$h) cluster operator is
\begin{equation}
T_k =
\frac{1}{(k!)^2} \sum_{i_1,\ldots,i_k; a_1,\ldots,a_k} t_{i_1\ldots
i_k}^{a_1\ldots a_k}
a^\dagger_{a_1}\ldots a^\dagger_{a_k}
a_{i_k}\ldots a_{i_1} \ .
\end{equation}
Here and in the following, the indices $i, j, k,\ldots$ label occupied single-particle 
orbitals
while $a,b,c,\ldots$ label unoccupied orbitals. The most commonly used
approximation is coupled-cluster with singles-and-doubles excitations
(CCSD) where $T\approx T_1+ T_2$. 
The unknown amplitudes $t_i^a$ and $t_{ij}^{ab}$
in Eq.~(\ref{T}) are determined from the solution of the coupled-cluster equations
\begin{eqnarray}
\label{ccsd}
0 &=& \langle \phi_i^a | \overline{H} | \phi_0\rangle \ , \\
0 &=& \langle \phi_{ij}^{ab} | \overline{H} | \phi_0\rangle \ .
\end{eqnarray}
Here $|\phi_i^a\rangle = a_{a}^\dagger a_{i}|\phi_0\rangle$
is a 1p-1h excitation of the reference state, and
$|\phi_{ij}^{ab}\rangle$ is a similarly defined 2p-2h excited
state. The CCSD equations~(\ref{ccsd}) thus demand that the reference
state $|\phi_0\rangle$ has no 1p-1h and no 2p-2h excitations, i.e., it is
an eigenstate of the similarity-transformed Hamiltonian~(\ref{hbar})
in the space of all 1p-1h and 2p-2h excited states. Once the CCSD
equations are solved, the ground-state energy is computed as 
\begin{equation}
E = \langle \phi_0 | \overline{H} | \phi_0\rangle \ .  
\end{equation}
Coupled-cluster
theory has the following virtues. First, the similarity-transformed
Hamiltonian can be evaluated exactly since the
Hausdorff-Baker-Campbell expansion 
\ba 
\overline{H} &=& H +
[H,T] + {1\over 2!} \left[ [H,T], T\right] \nonumber\\
&&+ {1\over 3!}\left[\left[ [H,T],T\right],T\right] + \ldots \nonumber\\
&=& \left(H e^T\right)_c 
\ea
terminates exactly at fourfold-nested commutators for two-body
Hamiltonians. Here, the last term in parentheses, $(\ldots)_c$, indicates that only those terms
enter where a cluster operator $T$ is connected to the Hamiltonian
$H$ to the right. Since every cluster operator is linked to the Hamiltonian, 
coupled-cluster theory by construction fulfills Goldstone's linked cluster
theorem and yields size-extensive results. This is particularly
important in applications to medium-mass nuclei. Second, within the
CCSD approximation, the computational effort scales as $n_o^2 n_u^4$
where $n_o$ and $n_u$ denote the number of occupied and unoccupied
orbitals in the reference state $|\phi_0\rangle$, respectively. Thus,
the computational effort is much smaller than within the configuration
interaction approach (or shell-model in nuclear physics) 
which exhibits a factorial scaling as function of the chosen single-particle space and
number of nucleons.

Coupled-cluster theory with inclusion of full triples (CCSDT)~\cite{ccsdt-n} is
usually considered to be too computationally expensive in most many-body 
systems of considerable size. Therefore triples corrections are 
usually taken into account perturbatively using the non-iterative CCSD(T)
approach described in Ref.~\cite{Deegan94}. Recently, a more
sophisticated way of including the full triples known as
the $\Lambda$-CCSD(T) approach, has been developed by 
Taube \emph{et al.} \cite{Kucharski98,Taube08}. In the
$\Lambda$-CCSD(T) approach the left-eigenvector solution of the CCSD
similarity-transformed Hamiltonian is utilized in the calculation of a
non-iterative triples correction to the coupled-cluster ground state
energy. The left eigenvalue problem is given by
\be
\label{left}
\langle\psi| \Lambda \overline{H} = E\langle\psi| \Lambda \ ,
\ee
were ${\Lambda}$ denotes the de-excitation cluster operator
\be
{\Lambda} = 1 + {\Lambda}_1 +{\Lambda}_2 \ ,
\ee
with
\ba
{\Lambda}_1 &=&
\sum_{i,a} \lambda^{i}_{a}
a_{a}
a^\dagger_{i} \ , \\
{\Lambda}_2 &=&
\frac{1}{4} \sum_{i,j,a,b} \lambda^{ij}_{ab}
a_{b} a_{a}
a^\dagger_{i} a^\dagger_{j} \ .
\ea
The unknowns, $\lambda^i_a$ and $\lambda^{ij}_{ab}$, result from the ground-state
solution of the left eigenvalue problem (\ref{left}). They are
utilized together with the cluster amplitudes, $t_i^a$ and
$t_{ij}^{ab}$, to compute the energy correction due to triples clusters
as
\ba
\Delta E_3&=& {1\over (3!)^2} \sum_{i j k a b c} \langle\psi|
{\Lambda}({F}_{hp} +{V})_N|\psi^{abc}_{ijk}\rangle\nonumber\\
&\times& {1\over
\varepsilon_{ijk}^{abc}}\langle\psi_{ijk}^{abc}|({V}_N{T}_2)_C|\psi\rangle
\ .  
\ea

Here, ${F}_{hp}$ denotes the part of the normal-ordered one-body
Hamiltonian that annihilates particles and creates holes, while 
\be
\varepsilon_{ijk}^{abc} \equiv f_{ii}+f_{jj}+f_{kk}-f_{aa}-f_{bb}-f_{cc}
\ee
is expressed in terms of the diagonal matrix elements of the
normal-ordered one-body Hamiltonian ${F}$. In the case of
Hartree-Fock orbitals, the one-body part of the Hamiltonian is
diagonal and ${F}_{hp}$ vanishes. The subscript $C$ denotes 
the connected part of the operator, and $|\psi_{ijk}^{abc}\rangle$ is a 3p-3h 
excitation of the reference state.

\subsection{Equation-of-motion coupled-cluster theory for ground and
  excited states of closed- and open-shell nuclei}
\label{subsec:eomccsd} 
The CCSD and the $\Lambda$-CCSD(T) approaches are known as
single-reference coupled-cluster methods (SR-CCM), and therefore they
are particularly well suited for nuclei with an expected closed (sub-)
shell structure.  By adding or removing particles to a closed-shell
nucleus, we move into regions of nuclei with open-shell structure, and
clearly the SR-CCM methods are not well suited for the description of
these nuclei.  Most nuclei are of open-shell character and in order to
study and predict properties like the evolution of shell-structure as
we move towards the drip line in various isotopic chains, we clearly
need to go beyond the SR-CCM class of approaches.  There exists a
variety of coupled-cluster methods that have been specifically
designed to address the structure of open-shell systems, see
Ref.~\cite{Bar07} for an overview of some of these methods. Most
methods are either based on a multi-reference formulation of the
coupled-cluster method (MR-CCM) (see Ref.~\cite{Bar02} for a
discussion of various SR-CCM and MR-CCM), or an extension of
equation-of-motion (EOM) theory \cite{Rowe68} based on the
coupled-cluster method (EOM-CCM) (see, e.g., Refs.~\cite{Bar84,
  Hirata00}).  The beauty of the EOM-CCM is that it has the simplicity
and transparency of SR-CCM such as the CCSD and the $\Lambda$-CCSD(T)
approaches discussed above, and that the method allows for systematic
improvements and extensions.

In order to extend our \emph{ab initio} coupled-cluster program beyond
closed (sub-) shell nuclei, we use the EOM-CCM approach. We give a
brief outline of the method in the following. The basic idea behind
EOM \cite{Rowe68} is to calculate states $\vert\psi_k\rangle $ of
the nucleus $B$ by acting with an excitation operator $\Omega_k$ on the
ground state $\vert \psi_0\rangle$ of a closed-shell reference nucleus
$A$.  If $\Omega_k$ is a particle number-conserving operator ($B=A$),
$\Omega_k$ will generate excited states of the nucleus $A$. In the case
of $\Omega_k$ not conserving the number of particles ($B \neq A $),
$\Omega_k$ can generate ground and excited states of the open-shell
nucleus $B$ with respect to the reference nucleus $A$. In EOM-CC theory,
the ground state of the closed-shell reference nucleus $A$ is given by
the coupled-cluster wave function, $ \vert \psi_0 \rangle = \exp(T)
\vert \phi_0 \rangle$.  In case we wish to calculate excited states of
the closed-shell nucleus $A$ ($B=A$), we define the excitation
operator $\Omega_k = R_k^A$,
\begin{equation} 
R_k^A = r_0 + \sum_{i,a}r^a_i a^\dagger _a a_i + {1\over 4} 
\sum_{i,j,a,b}r^{ab}_{ij} a^\dagger _a a^\dagger_b a_ja_i + \ldots  \ ,
\label{rex}
\end{equation}
and by truncating at the two-particle-two-hole excitation level, we get the 
standard EOM-CCSD approach. In case we wish to calculate ground 
and excited states of a nucleus with a nucleon added or removed from 
a closed-shell nucleus ($B=A\pm1$), we define the excitation operators
$\Omega_k = R_k^{A\pm 1}$ with,  
\begin{eqnarray}
R_k^{(A+1)} = \sum_a r^aa_a^{\dagger} + {1\over
  2} \sum_{j,a,b}r^{ab}_j a^{\dagger}_a  a^{\dagger}_b  a_j + \ldots \ , \nonumber\\
R_k^{(A-1)} = \sum_i r_ia_i + {1\over
  2} \sum_{i,j,a}r^{a}_{ij} a^{\dagger}_a  a_i  a_j + \ldots \ ,
\label{eq:paeqn1}
\end{eqnarray} 
and truncating at the two-particle-one-hole and two-hole-one-particle
levels, we get the the particle-attached and particle-removed
equation-of-motion coupled-cluster methods (PA/PR-EOM-CCM), see, for
example, Refs. \cite{Hirata00,Bar07}.  The PA-EOM-CCM adds a particle
to a nucleus with $A$ nucleons by creating $1p$, $2p$-$1h$,
$3p$-$2h,\ldots$, excitations on the ground state of the nucleus
with $A$ nucleons. Similarly, PR-EOM-CCM removes a particle from $A$
by creating $1h$, $1p$-$2h$, $2p$-$3h, \ldots$, excitations on the
ground state of $A$. By multiplying the equation for the nucleus $A$,
$\overline{H} \vert\phi_0\rangle = E_0\vert \phi_0\rangle $, from the
left with $\Omega_k$, and subtracting it from the equation for the 
nucleus $B$, $\overline{H} \Omega_k\vert\phi_0\rangle = E_k
\Omega_k\vert\phi_0\rangle $, we obtain the following equation,
\begin{equation}
\left[ \overline{H}, \Omega_k \right] \vert \phi_0 \rangle = 
\left( \overline{H} \Omega_k \right)_C \vert\phi_0\rangle = \omega_k
\Omega_k \vert \phi_0 \rangle \ ,
\label{eq:paeqn2}
\end{equation}
with $\overline{H}=e^{-{T}} {H}e^{{T}}$ being the similarity-transformed 
Hamiltonian and $\omega_k = E_k-E_0$ the energy of the nucleus
$B$ relative to the ground state of the nucleus $A$.
Eq.~(\ref{eq:paeqn2}) defines a right eigenvalue problem for the
excitation amplitudes in Eqs.~(\ref{rex}) and (\ref{eq:paeqn1}), and
is usually solved using iterative eigenvalue algorithms.

\subsection{Spherical coupled-cluster theory} 
For ``hard'' interactions such as the ``bare'' N$^3$LO interaction, wave
function methods need to employ very large model spaces in order to
yield converged results. For spherical reference states (nuclei
with closed major shells or closed subshells), it is therefore better
to employ the spherical symmetry to further reduce the number of
unknowns, that is, the number of cluster amplitudes. For such nuclei, the
cluster operator of Eq.~(\ref{T}) is a scalar under rotation, and depends
only on reduced amplitudes. Thus,
\begin{equation}
{T}_1= \sum_{j_i j_a} t_{j_i}^{j^a}(a_{j_a}^\dagger \times
\tilde{a}_{j_i})^{(0)} \ ,
\end{equation}
and 
\begin{equation}
{T}_2= \sum_{j_i j_j j_a j_b J} t_{j_i j_j}^{j_a j_b}(J)
(a_{j_a}^\dagger \times a_{j_b}^\dagger)^{(J)}\cdot
(\tilde{a}_{j_j}\times \tilde{a}_{j_i})^{(J)} \ .
\end{equation}
Here, we employed the usual notation for spherical tensors, and $j_i$
and $j_a$ denote the spin of the occupied and unoccupied subshells,
respectively. It is clear that the similarity-transformed Hamiltonian
is also a scalar under rotation, and it is straightforward to work out the 
CCSD equations within this
formulation. Details are given in the Appendix.
The computational cost of PA-EOM-CCSD is like that of CCSD, i.e., 
$n_u^4n_o^2$, where $n_u$ is the number of
unoccupied orbitals and $n_o$ is the number of occupied orbitals.  
In order to reduce the memory and computational cost related to the 
basis size, we write the 
excitation amplitudes $R_k^{(A\pm1)}$ as spherical tensors of rank $J$ and
projection$M$,  
\begin{eqnarray}
\lefteqn{R_k^{(A+1)}(J,M) = \sum_{j_a}r^{j_a}(J) a_{j_a m_a}^{\dagger}
\delta_{J}^{j_a}\delta_{M}^{m_a}}  \\ 
& & + {1\over2} \sum_{j_j j_a j_b J_{ab}}r^{j_a j_b}_{j_j}(J_{ab}, J) \left[
\left( a^{\dagger}_{j_a} \otimes a^{\dagger}_{j_b}\right) ^{J_{ab}}
\otimes \tilde{a}_{j_j}\right] ^{(J)}_M \ , \nonumber\\
\lefteqn{R_k^{(A-1)}(J,M) = \sum_{j_i}r^{j_i}(J) a_{j_i m_i}
\delta_{J}^{j_i}\delta_{M}^{m_i} }  \\ 
& & + {1\over2} \sum_{j_i j_j j_a J_{ij}} r^{j_a}_{j_i j_j}(J_{ij},J) \left[
{a}^\dagger_{j_a} \otimes \left( \tilde{a}_{j_i} \otimes \tilde{a}_{j_j}\right) ^{J_{ij}}
\right] ^{(J)}_M \ .\nonumber
\label{eq:paeqn3}
\end{eqnarray}
Here, we wish to solve for the reduced excitation amplitudes 
$r^{j_a}(J)$ and $r^{j_a,j_b}_{j_j}(J_{ab},J)$. In this coupled
scheme, the eigenvalue equation in Eq.~(\ref{eq:paeqn2}) 
is solved separately for each set of quantum numbers $\{J^{\pi}, T_z\}$.

A simple estimate shows that a model space of $n_o+n_u$
single-particle states consists of only $(n_o+n_u)^{2/3}$ $j$-shells.
Thus, the entire computational effort is approximately reduced by a
power $2/3$ within the spherical scheme compared to the $m$-scheme. We
have derived and numerically implemented the spherical scheme within
the CCSD approximation. We tested that our $m$-scheme code and the
spherical code give identical results for several test cases. The
spherical code permits us to reach much larger model spaces, and we
are able to achieve satisfactory convergence even for ``hard''
interactions such as the ``bare'' N$^3$LO interaction. In
Ref.~\cite{Hag08} we used the spherical coupled-cluster code to
calculate the ground states of medium-mass nuclei like $^{40,48}$Ca
and $^{48}$Ni within the CCSD approximation and starting from bare
chiral interactions. In a model space of 15 major oscillator shells,
the results were reasonably well converged. Recently, we have also
implemented the $\Lambda$-CCSD(T) approach in an
angular-momentum-coupled scheme. In Ref.~\cite{Hag09_ox}, we
calculated the ground states of the oxygen isotopes using chiral
interactions in model spaces comprising up to 20 major oscillator
shells within the $\Lambda$-CCSD(T) approach.  In such large spaces we
were even able to converge the ground state energies for a chiral
interaction with a 600~MeV$c^{-1}$ momentum cutoff. Likewise, the
inclusion of continuum scattering states -- necessary for a
description of halo states in weakly bound nuclei -- yields large
model spaces that can be treated within he spherical
scheme~\cite{hagen10}. Details of the angular-momentum-coupled
$\Lambda$-CCSD(T) approach are given in the Appendix.

\subsection{Expectation values in coupled-cluster theory}
For expectation values and the computation of ground state properties
other than the energy, we utilize the right and left eigenvectors of the
similarity-transformed Hamiltonian $\overline{H}$ and compute the one-
and two-body reduced density matrices,
\begin{eqnarray} 
\rho_{pq} = \langle\phi_0\vert \Lambda e^{-T}a^\dagger_pa_q e^T \vert \phi_0\rangle \ , \\
\rho_{pqrs} = \langle\phi_0\vert \Lambda e^{-T}a^\dagger_pa^\dagger_qa_ra_s e^T \vert \phi_0\rangle \ .
\end{eqnarray}
Expectation values can then be computed by expressing the operator of
interest in terms of the density matrices.  The Hellmann-Feynman
theorem is another route to the computation of expectation values. In
this case, we consider a response of the ground-state energy to a
perturbation caused by a given operator of interest. The
Hellmann-Feynman theorem expresses the expectation values of an
observable $B$ as
\begin{equation}
\label{expect}
\langle B\rangle = {\partial
  E(\beta)\over\partial\beta}\bigg|_\beta=0 \ .  
\end{equation}
Here $E(\beta)$ is the ground-state energy of the Hamiltonian $H+\beta
B$.  Unfortunately, the relation~(\ref{expect}) does not hold exactly
within the coupled-cluster approach since this method is not
variational when the cluster operator is truncated. We will
nevertheless base the computation of expectation values on
Eq.~(\ref{expect}).  Experience shows that this approach is
approximately valid if the relevant particle-hole excitations are
incorporated. In practice, the differential quotient~(\ref{expect}) is
numerically implemented as a difference quotient with $\beta\approx
0.01$. This requires us to perform two calculations (one for zero and
one for small nonzero $\beta$). For relatively ``hard'' interactions
such as the ``bare'' N$^3$LO, we also perform two Hartree-Fock
calculations and thus employ slightly different single-particle bases
in these two cases. While such a procedure is unnecessary for methods
that fulfill the Hellmann-Feynman theorem, it is important within our
approach. We note that this approach most closely reflects the
physical situation of an expectation value measuring the system's
response to a perturbation

\section{Treatment of the center-of-mass problem} 
\label{sec:com}
In this section, we demonstrate that the coupled-cluster wave function
factorizes to a very good approximation into a product of a
center-of-mass wave function and an intrinsic wave function. We will
present a simple procedure that checks this factorization and give an
estimate for the degree of the achieved factorization. This section
significantly expands on the short demonstration of the factorization in
Ref.~\cite{Hagen_PRL2009}.

\subsection{Statement of the center-of-mass problem}
Let us consider the nuclear $A$-body Hamiltonian
\begin{equation}
\label{hamA}
{H}_A=\sum_{j=1}^A {\vec{p}^2\over 2m} +\sum_{j<k}^A {V}(j,k) \ .
\end{equation}
Here, ${V}$ is a two-body operator that is invariant under
rotations and translations.  Thus, the total momentum and angular
momentum are conserved quantities. It is advantageous to separate the
Hamiltonian into an intrinsic Hamiltonian ${H}_{\rm in}$ and the
center-of-mass Hamiltonian (i.e., the kinetic energy ${T}_{\rm cm}$
of the center of mass) as
\ba
{H} &=& {T}_{\rm cm} +{H}_{\rm in} \nonumber\\
&=& {T}_{\rm cm} +\sum_{j < k}^A\left( {(\vec{p}_j-\vec{p}_k)^2\over 2mA} 
+ {V}(j,k)\right)  \ .
\ea
Note that the intrinsic Hamiltonian ${H}_{\rm in}$ does not depend
on the center-of-mass coordinate. Thus, we could add an arbitrary
operator ${H}_{\rm cm}$ of the center-of-mass coordinate to the
intrinsic Hamiltonian ${H}_{\rm in}$ without changing the
intrinsic properties of the resulting Hamiltonian. In other words, the
eigenfunctions $\psi_A$ of the $A$-body Hamiltonian ${H}_{\rm
  in}+{H}_{\rm cm}$ are products of an intrinsic eigenfunction
$\psi_{\rm in}$ of the intrinsic Hamiltonian ${H}_{\rm in}$, and a
center-of-mass wave function $\psi_{\rm cm}$ that is the eigenfunction
of the center-of-mass Hamiltonian ${H}_{\rm cm}$, that is, 
\begin{equation}
\label{factor}
\psi_A=\psi_{\rm cm}\psi_{\rm in}  \ .
\end{equation}
The corresponding energy is the sum $E_A=E_{\rm cm} + E_{\rm in}$ of
the center-of-mass energy and the intrinsic energy. The question thus
arises which operator ${H}_{\rm cm}$ to choose. Let us consider the
NCSM~\cite{NCSM_Review} as an example.  Here, one works in a complete
$N\hbar\omega$ space consisting of a basis of all $A$-particle Slater
determinants of oscillator states with frequency $\omega$ and total
excitation energy not exceeding $N\hbar\omega$. One can add the
center-of-mass Hamiltonian
\begin{equation}
\label{Hosc}
{H}_{\rm cm}(\tilde{\omega}) = {T}_{\rm cm} + 
{1\over 2}mA\tilde{\omega}^2 R_{\rm cm}^2 -{3\over 2} \hbar\tilde{\omega}
\end{equation}
with $\tilde{\omega}=\omega$ to the intrinsic Hamiltonian
${H}_{\rm in}$, and obtains a factorized ground-state wave
function where the center-of-mass wave function $\psi_{\rm cm}$ is a
Gaussian with frequency $\omega$. The (truncated) coupled-cluster
method is unable to employ an $N\hbar\omega$ space, and it is arguably
the best idea to completely remove any reference to the center-of-mass
coordinate. Thus, we solve an intrinsic Hamiltonian ${H}_{\rm in}$
that depends on $A-1$ independent coordinates in a Hilbert space
spanned by wave functions of $A$ coordinates. Two comments are in
order. First, the truncated coupled-cluster method is not an exact
solution of the $A$-body problem, but rather a very efficient
approximation. Thus, it is not guaranteed {\it a priori} that the
coupled-cluster wave function exhibits the
factorization~(\ref{factor}).  Recall that any $A$-body wave function
can be expanded as
\begin{equation}
\label{svd}
\psi_A=\sum_{j\ge 1} s_j \psi_{\rm cm}^{(j)}\psi_{\rm in}^{(j)}  \ ,
\end{equation}
where $\sum_j s_j^2 = 1$ from normalization, and we assume that the
non-negative weights $s_j$ are ordered in decreasing order.  Only if
all but one of the weights $s_j$ vanishes, does the
factorization~(\ref{factor}) take place; otherwise, the factorization
might be mildly or strongly violated, depending on the size of the
weights $s_j$. Second, if a factorization of the coupled-cluster wave
function takes (approximately) place, the form of the center-of-mass
wave function $\psi_{\rm cm}^{(1)}$ corresponding to the largest
weight $s_1$ has to be determined.  In what follows we will see that
the $A$-body coupled-cluster wave function factorizes to a good
approximation (i.e., $s_1\approx 1$), and that the center-of-mass wave
function is the Gaussian ground state of the center-of-mass
Hamiltonian (\ref{Hosc}) for a yet-to-be determined frequency
$\tilde{\omega}$.

\subsection{Approximate separation of the center-of-mass wave function
  in coupled-cluster calculations}

\begin{figure}[h]
\includegraphics[width=0.45\textwidth,clip=]{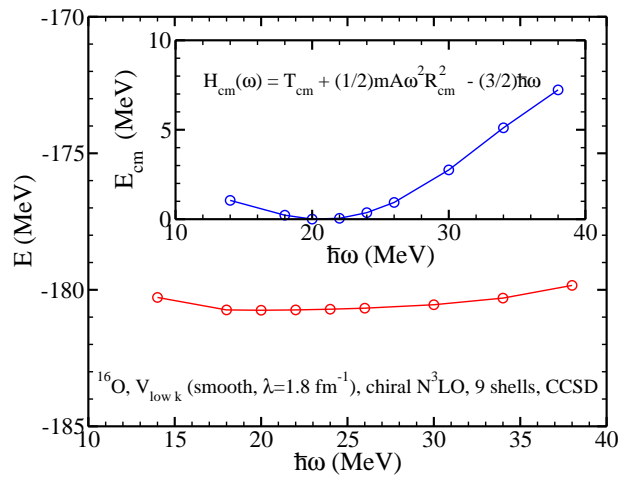}
\caption{(Color online) Ground-state energy (within CCSD) of
$^{16}$O with a low-momentum interaction as a function of the oscillator
spacing $\hbar\omega$. The model space consists of nine major
oscillator shells. Inset: Expectation value $E_{\rm cm}(\omega)$ of the center-of
mass Hamiltonian with the standard frequency dependence.} 
\label{fig:fig1}
\end{figure}

\begin{figure}[h]
\includegraphics[width=0.45\textwidth]{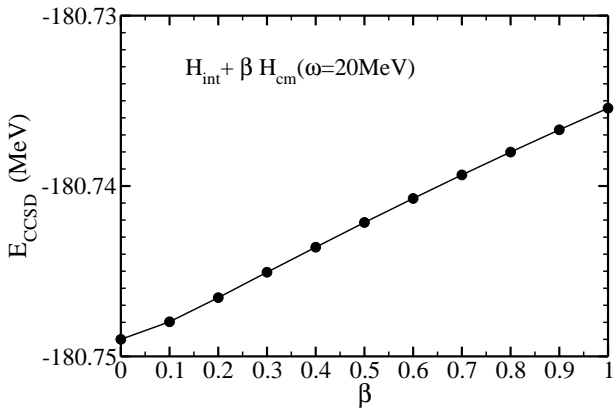}
\caption{(Color online) Ground-state energy (within CCSD) of $^{16}$O
  using a low-momentum interaction with the center-of-mass Hamiltonian
  $\beta H_{\rm cm}$ added. Calculations employ nine major oscillator
  shells and $\hbar\omega = 20$~MeV.  As $\beta$ is varied between 0
  and 1, the ground state energy changes by about $15$~keV.}
\label{fig:fig2}
\end{figure}

We consider the nucleus $^{16}$O and employ the low-momentum
interaction $\vlowk$ with a smooth momentum cutoff
$\lambda=1.8$~fm$^{-1}$ derived from Machleidt's and Entem's chiral
N$^3$LO interaction. Figure~\ref{fig:fig1} shows the ground-state energy
as a function of the oscillator spacing $\hbar\omega$ of the
underlying oscillator basis in a model space of nine oscillator
shells. These results are obtained from a CCSD calculation within a
spherical Hartree-Fock basis. The energy is well converged as it
depends very weakly on the model-space parameters. We also computed
the ground-state expectation value
\begin{equation}
\label{ecm}
E_{\rm cm}(\tilde\omega) = \langle {H}(\tilde{\omega})\rangle 
\end{equation}
of the center-of-mass Hamiltonian~(\ref{Hosc}) for a frequency
$\tilde\omega=\omega$ and show the result in the inset of
Fig.~\ref{fig:fig1}. This expectation value is generally not zero. This
indicates that the coupled-cluster wave function is in general not an
eigenstate of the center-of-mass Hamiltonian ${H}(\omega)$.
However, there seems to be little correlation between the
ground-state energy and the expectation value $E(\omega)$, and the
latter vanishes approximately in a model space with
$\hbar\omega\approx 20$~MeV. Thus, at this frequency, the
coupled-cluster wave function is approximately the ground state of the
center-of-mass Hamiltonian ${H}(\omega)$. As a check, we fix
$\hbar\omega=20$~MeV, consider the Hamiltonian ${H}={H}_{\rm
  in}+\beta{H}(\omega)$, and compute its ground-state energy as a
function of the parameter $\beta$. The result is shown in
Fig.~\ref{fig:fig2}.  Clearly, the ground-state energy is rather
insensitive to $\beta$ and varies by only 15~keV as $\beta$ is
increased from zero to one. 

\begin{figure}[t]
  \includegraphics[width=0.45\textwidth]{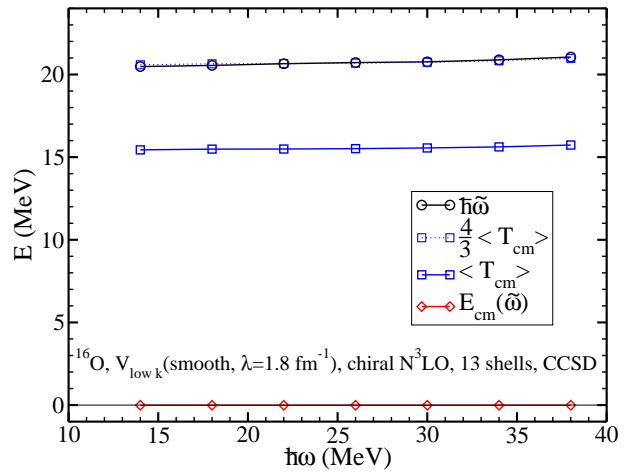}
  \caption{(Color online) 
    $^{16}$O Ground-state expectation value (within CCSD) of the generalized center-of-mass
    Hamiltonian $H_{\rm cm}(\tilde{\omega})$ and of the kinetic energy
    $T_{\rm cm}$ as a function of the
    oscillator spacing $\hbar\omega$. 
    The model space consists of
    thirteen major oscillator shells.}
  \label{fig:fig3}
\end{figure}

The factorization of Eq.~(\ref{factor}) is thus achieved in a model space with 
$\hbar\omega\approx 20$~MeV. Let us assume that such a factorization
takes place generally, and that the corresponding center-of-mass wave
function is a Gaussian with unknown frequency $\tilde{\omega}$
that might differ from the frequency of the underlying oscillator
basis. Thus, we assume that the coupled-cluster wave function is the
ground state of the center-of-mass Hamiltonian~(\ref{Hosc}) for a
suitable frequency $\tilde\omega$. To determine this frequency we
employ the identity 
\ba 
{H}_{\rm cm}(\omega)+{3\over
  2}\hbar\omega-{T}_{\rm cm} =
{\omega^2\over\tilde{\omega}^2}\left({H}_{\rm cm}(\tilde{\omega})
+{3\over 2}\hbar\tilde{\omega}-{T}_{\rm cm}\right) \ ,\nonumber
\ea 
and take its expectation value. We seek (and insert) $E_{\rm
  cm}(\tilde{\omega})=0$, employ the relation $\langle {T}_{\rm
  cm}\rangle={3\over 4}\hbar\tilde{\omega}$ valid for Gaussians,
insert the already computed expectation values $E_{\rm cm}(\omega)$,
and solve for the unknown frequency $\tilde{\omega}$. This yields the
two possible solutions.
\ba
\label{magic}
\hbar\tilde{\omega} &=& \hbar\omega + {2\over 3} E_{\rm cm}(\omega) \nonumber\\
&& \pm \sqrt{{4\over 9}(E_{\rm cm}(\omega))^2 +{4\over 3}\hbar\omega E_{\rm
cm}(\omega)} \ .  
\ea 
We compute the ground-state expectation values $E_{\rm
  cm}(\tilde\omega)$ for these two frequencies and find that one
expectation value is typically very close to zero. Figure~\ref{fig:fig3}
shows that the small expectation value essentially vanishes for a
large range of frequencies of the underlying oscillator basis in a
model space of 13 major oscillator shells. Closer inspection shows
that the expectation value is about $E_{\rm cm}(\tilde\omega)\approx
-10$~keV. Recall that coupled-cluster theory is non-variational (as the
similarity-transformed Hamiltonian is non-Hermitian), and such a small
negative expectation value is certainly tolerable for the non-negative
operator~(\ref{Hosc}). Figure~\ref{fig:fig3} also shows that the frequency
of the Gaussian center-of-mass wave function stays approximately
constant $\hbar\tilde\omega\approx 20$~MeV for a large range of frequencies
$\omega$ of the underlying oscillator basis. As a final check, we also
computed the ground-state expectation value of the center-of-mass
kinetic energy. For a Gaussian, this expectation value fulfills
$\langle{T}_{\rm cm}\rangle = {3\over
  4}\hbar\tilde\omega$. Figure~\ref{fig:fig3} shows that this relation is
indeed obeyed.

\begin{figure}[t]
  \includegraphics[width=0.45\textwidth,clip=]{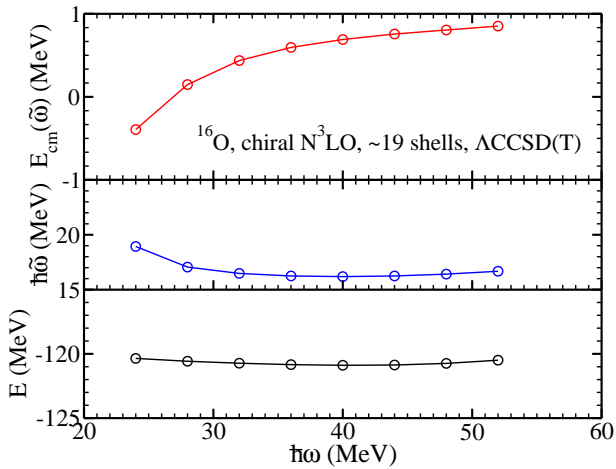}
  \caption{(Color online) 
    Bottom: Ground-state energy of $^{16}$O within
    the $\Lambda$-CCSD(T) approximation as a function of the frequency
    $\hbar\omega$ of the underlying oscillator basis. Middle: Relation
    between the frequency $\tilde{\omega}$ and the frequency $\omega$ of
    the underlying oscillator basis. Top: Expectation value $E_{\rm
      cm}(\tilde{\omega})$ of the center-of-mass vs. $\hbar\omega$.}
  \label{fig:fig4}
\end{figure}

\begin{figure}[t]
  \includegraphics[width=0.45\textwidth,clip=]{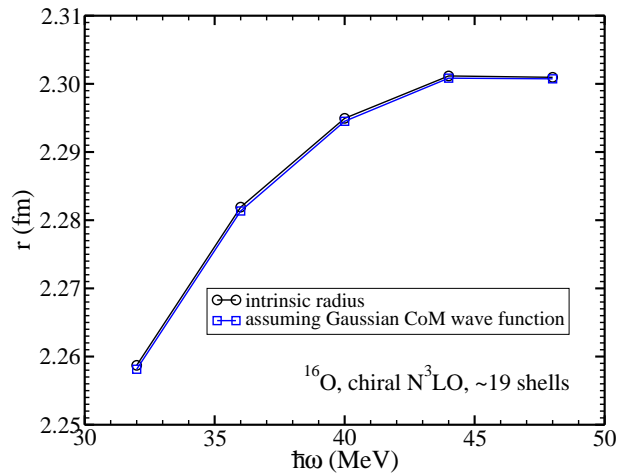}
  \caption{(Color online) RMS radii of $^{16}$O within the 
    $\Lambda$-CCSD(T) approximation using the intrinsic point radius and 
    subtracting $\langle R_{\rm cm}^2 \rangle$ 
    assuming a Gaussian for the center of mass.}
  \label{fig:fig5}
\end{figure}

Let us turn to the ``bare'' N$^3$LO interaction \cite{machleidt02,N3LO}. This
interaction has a substantially higher momentum cutoff than the
previously employed low-momentum interaction. For converged results, we
need to employ larger model spaces and also triples clusters. We
performed coupled-cluster calculations of the ground states of
$^{16}$O and $^{4}$He. Figure~\ref{fig:fig4} shows the ground-state energy
of $^{16}$O (lower panel), the ground-state expectation value $E_{\rm
  cm}(\tilde\omega)$ (upper panel) of the center-of-mass Hamiltonian~(\ref{Hosc}),
and the corresponding frequency $\tilde\omega$ (middle panel) as a
function of the frequency $\omega$ of the underlying oscillator basis.
One sees that the ground-state energy is well converged and displays
only a weak dependence on $\omega$. Similar comments apply to the
frequency $\tilde\omega$. The expectation value $E_{\rm
  cm}(\tilde\omega)$ is small, i.e., $|E_{\rm cm}(\tilde\omega)|\ll
\hbar\tilde\omega$ but nonzero. For smaller values of the frequency
$\omega$ of the underlying oscillator basis, we are at the limit of
well-converged results, and $E_{\rm cm}$ even becomes negative. This
shows that the coupled-cluster ground state is not an exact eigenstate of
the center-of-mass Hamiltonian~(\ref{Hosc}). Let us also assume that
the factorization is not perfect (though this is not implied by a
nonzero value of $E_{\rm cm}$). We want to estimate the level of
admixture of center-of-mass excitations. A simple and conservative
estimate based on perturbation theory shows that the admixture is
essentially $E_{\rm cm}/\hbar\tilde\omega\approx 5\%$.
We also computed the intrinsic point radius $r$ from 
\begin{equation}
\label{rad1}
r^2\equiv {1\over A^2}\sum_{1\le j<k\le A} \langle (\vec{r}_j-\vec{r}_k)^2\rangle
\end{equation}
and show the result in Fig.~\ref{fig:fig5}. This calculation is again
based on the triples-corrected $\Lambda$-CCSD(T) approximation, and it still
exhibits a weak model-space dependence. It provides us with a further
test of the wave-function factorization, since we can employ the
center-of-mass coordinate $\vec{R}_{\rm cm}$ and rewrite
Eq.~(\ref{rad1}) as
\begin{equation}
\label{rad2}
r^2 = {1\over A}\sum_{j=1}^A\langle (\vec{r}_j-\vec{R}_{\rm cm})^2\rangle = 
{1\over A}\sum_{j=1}^A\langle \vec{r}_j^2 \rangle - \langle\vec{R}_{\rm cm}^2\rangle \ . 
\end{equation}
For a Gaussian corresponding to the frequency $\tilde\omega$, we have
${1\over 2}mA\tilde\omega^2\langle\vec{R}_{\rm cm}^2\rangle = {3 \over
  4}\hbar\tilde\omega$, and the intrinsic point radius can thus be computed
from the expectation value of a one-body operator. The result from
this calculation is in very good agreement with the result obtained
from Eq.~(\ref{rad1}), as shown in Fig.~\ref{fig:fig5}. This suggests that
the factorization of the wave function might be better than expected
from the calculation of the ground-state expectation value~(\ref{ecm})
depicted in Fig.~\ref{fig:fig4}.

At the moment, we have no analytical insights into the observed
factorization and the Gaussian shape of the center-of-mass wave
function. We believe that the factorization itself is not entirely
surprising. While our finite basis is not complete in a mathematical
sense, it is sufficiently complete to describe low-energy nuclear
structure, i.e., it contains momentum modes that exceed the cutoff of
the interaction, and it is sufficiently extended in position space to
accommodate a quantum object of the size of the nucleus. Thus, one
expects the basis to capture the relevant physics, and a
well-factorized ground state is not too surprising, even if the
employed many-body basis functions do not individually reflect this
factorization. 

\subsection{Center-of-mass problem for open-shell nuclei}
\begin{figure}[t]
  \includegraphics[width=0.45\textwidth,clip=]{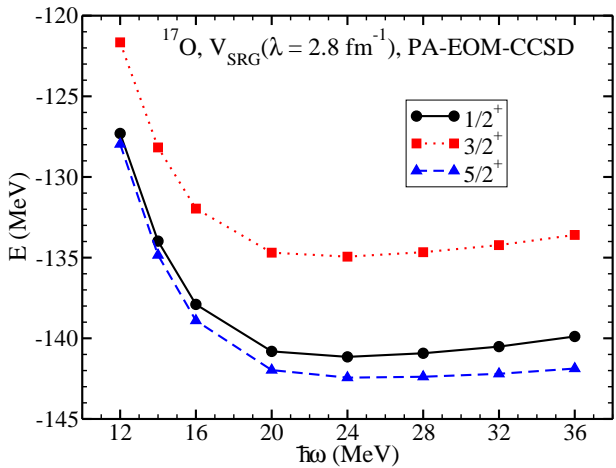}
  \caption{(Color online) Energies of low-lying $J^\pi=1/2^+$,
    $J^\pi=3/2^+$, $J^\pi=5/2^+$ states in $^{17}$O based on an SRG
    interaction with cutoff $\lambda=2.8$~fm$^{-1}$ in a model space
    consisting of $N+1=13$ oscillator shells versus the oscillator
    frequency.}
  \label{fig:o17erg}
\end{figure}

\begin{figure}[t]
  \includegraphics[width=0.45\textwidth,clip=]{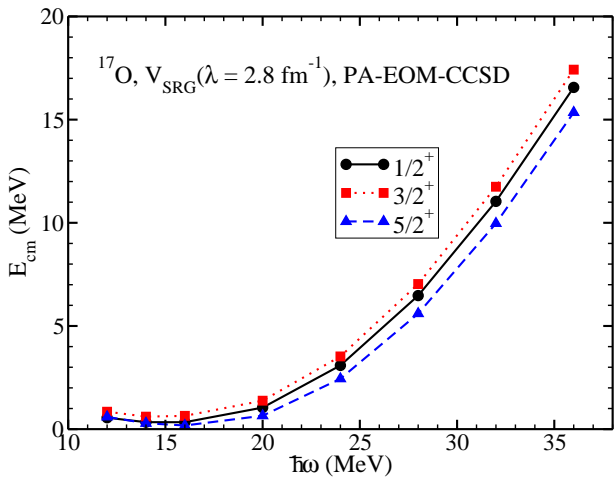}
  \caption{(Color online) Expectation values $E_{\rm cm}(\omega)$ of
    the center-of-mass Hamiltonian $H_{\rm cm}(\omega)$ (see
    Eq.~(\ref{Hosc})) for the low-lying $J^\pi=1/2^+$, $J^\pi=3/2^+$,
    $J^\pi=5/2^+$ states in $^{17}$O versus the oscillator frequency.}
  \label{fig:o17ecm_omega}
\end{figure}

Odd-mass nuclei that differ from a nucleus with closed subshells by
the addition or removal of one nucleon can also be computed
efficiently within the coupled-cluster method. For the calculation of
$^{17}$O, we employ the particle-attached equation-of-motion
method~(\ref{eq:paeqn2}), which is based on describing $^{17}$O as a
nucleon (i.e a superposition of 1p and 2p-1h excitations) upon the
ground state of $^{16}$O.  We first solve the coupled-cluster
equations~(\ref{ccsd}) for the ``mass-shifted'' nucleus $^{16}$O, i.e.,
for the computation of the $^{16}$O ground state, we employ the
Hamiltonian~(\ref{hamA}) with mass number $A=17$.  In a second step,
we describe the ground and excited states in $^{17}$O in terms of the
excitations~(\ref{eq:paeqn1}) upon this $^{16}$O ground state. For an
SRG interaction with cutoff $\lambda=2.8$~fm$^{-1}$, the energies of a
few low-lying states are shown in Fig.~\ref{fig:o17erg} as a function
of the oscillator frequency in a model space with $N=12$. These states
are single-particle states and can be well computed within our
approach.  Figure~\ref{fig:o17ecm_omega} shows the energy expectation
value $E_{\rm cm}(\omega)$ (defined in Eq.~(\ref{ecm})) for the three
low-lying single-particle states as a function of the oscillator
frequency $\omega$ of the underlying model space. These expectation
values are small for $\omega\approx 16$~MeV, but large for other
parameters. This shows that the center-of-mass wave function is
generally not a Gaussian with frequency $\omega$. No further
conclusion can be reached from this expectation value. In particular,
this is no evidence that the center-of-mass wave function does not
separate.

\begin{figure}[t]
  \includegraphics[width=0.45\textwidth,clip=]{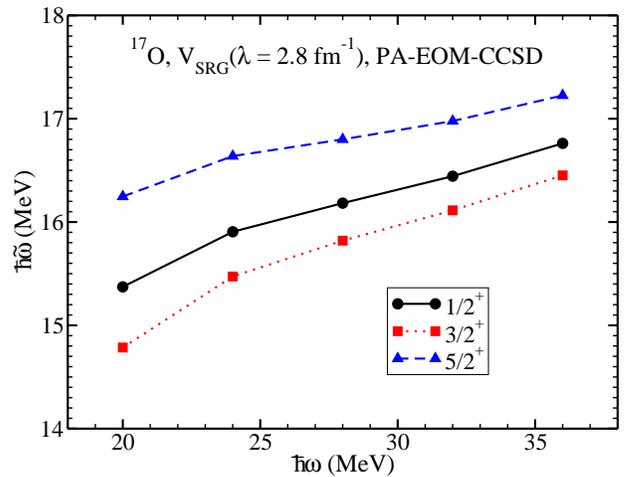}
  \caption{(Color online) Frequency $\tilde\omega$ of the
    (approximately) Gaussian center-of-mass wave function for the
    low-lying $J^\pi=1/2^+$, $J^\pi=3/2^+$, $J^\pi=5/2^+$ states in
    $^{17}$O versus the oscillator frequency.}
  \label{fig:o17omegatilde}
\end{figure}

\begin{figure}[t]
  \includegraphics[width=0.45\textwidth,clip=]{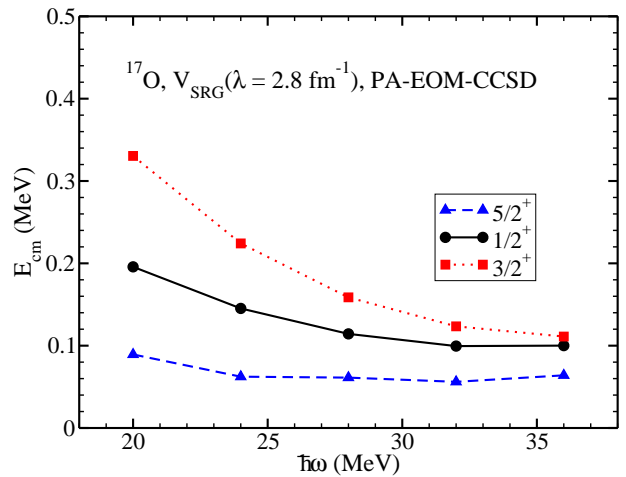}
  \caption{(Color online) Expectation values $E_{\rm cm}(\tilde\omega)$
    of the center-of-mass Hamiltonian $H_{\rm cm}(\tilde\omega)$ (see
    Eq.~(\ref{Hosc})) for the low-lying $J^\pi=1/2^+$, $J^\pi=3/2^+$,
    $J^\pi=5/2^+$ states in $^{17}$O versus the oscillator frequency.
    Here, the frequency $\tilde\omega=\tilde\omega(\omega)$ employed
    in the center-of-mass Hamiltonian is as depicted in
    Fig.~\ref{fig:o17omegatilde}. We have $E_{\rm cm}(\tilde\omega)\ll
    \hbar\tilde\omega$ and thus a very good separation of
    center-of-mass excitations.}
  \label{fig:o17ecmtilde}
\end{figure}

Let us again assume that the coupled-cluster wave function factorizes
into a Gaussian (with frequency $\tilde\omega$) for the center of mass
and an intrinsic wave function, and let us determine $\tilde\omega$
from Eq.~(\ref{magic}). Figure~\ref{fig:o17omegatilde} shows the
resulting frequency for the three states we computed in $^{17}$O. The
corresponding expectation values $E_{\rm cm}(\tilde\omega)$ defined in
Eq.~(\ref{ecm}) are shown in Fig.~\ref{fig:o17ecmtilde} as a function
of the oscillator frequency $\omega$ of the underlying model space.
We employ $\tilde\omega=\tilde\omega(\omega)$ as depicted in
Fig.~\ref{fig:o17omegatilde}. As evident from
Figs.~\ref{fig:o17omegatilde} and \ref{fig:o17ecmtilde}, we have
$E_{\rm cm}(\tilde\omega)\ll\hbar\tilde\omega$.  Thus, the computed
states in $^{17}$O also exhibit a Gaussian center-of-mass wave
function to a very good approximation (admixtures of higher
$\hbar\tilde\omega$ oscillator excitations are of the order of $E_{\rm
  cm}(\tilde\omega)/\hbar\tilde\omega\approx$~1--2\%). This implies
that the factorization is at least of the same degree of quality.

\subsection{Center-of-mass problem in a toy model}
\begin{figure}[t]
  \includegraphics[width=0.45\textwidth,clip=]{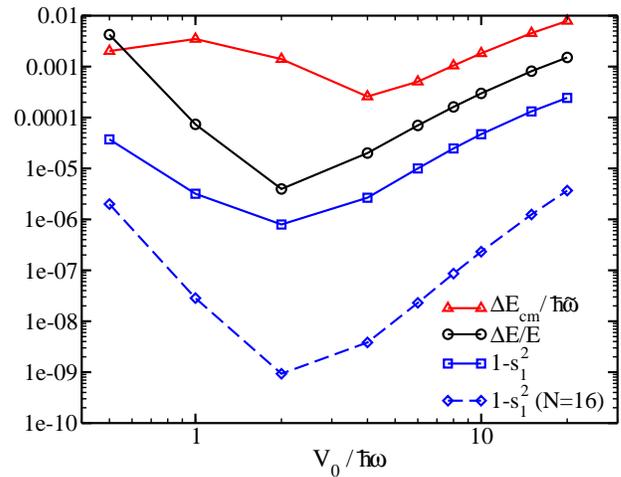}
  \caption{(Color online)Results for the toy model~(\ref{toy}) as a
    function of the strength of the interaction $V_0$. Circles: Estimate
    of relative error $\Delta E/E$ of the ground-state energy by
    comparing a calculation in $N=8$ oscillator shells with a
    calculation in $N=16$ shells. Squares and diamonds: Fraction
    $1-s_1^2$ of the ground state wave function that is not factorized
    in an intrinsic state and a center-of-mass state, obtained from
    the singular value decomposition~(\ref{svd}) in a model space of
    $N=8$ and $N=16$ oscillator shells, respectively. Triangles: The
    ground-state expectation value $E_{\rm cm}(\tilde\omega)$ of the
    harmonic oscillator center-of-mass Hamiltonian is much smaller
    than the energy $\hbar\omega$ of spurious center-of-mass
    excitations, indicating that the center-of-mass wave function is
    approximately a Gaussian. }
  \label{fig:afterfig5}
\end{figure}

To shed some light onto the observed factorization, we consider a
simple problem of two interacting particles in one spatial dimension.
We choose a two-body interaction of the form $V(x)=-V_0 \exp{(- (x/l)^2)}$,
where $x=(x_1-x_2)/\sqrt{2}$ is (up to a factor $\sqrt{2}$) the
relative coordinate of the two-particle system, and $l$ is a length
scale. We thus consider the intrinsic Hamiltonian 
\be
\label{toy}
H={p^2\over 2m}-V_0 \exp{(- (x/l)^2)}
\ee
with $m$ being the mass and $p$ the relative momentum
$p=(p_1-p_2)/\sqrt{2}$.  We choose a basis consisting of products
$\Phi_m(x_1/l)\Phi_n(x_2/l)$ of oscillator wave functions $\Phi_k$,
and choose $0\le m,n\le N$. Thus, the basis is not a complete
$N\hbar\omega$ space, and we solve an intrinsic Hamiltonian that
depends on the relative coordinate in a model space consisting of
single-particle coordinates. We consider different interaction
parameters $V_0$, and set the oscillator length of our basis equal to
the scale $l$ of the Gaussian interaction. Note that the resulting
ground state has an extension that, depending on the strength $V_0$ of
the interaction, differs considerably from $l$. Indeed, approximating
the interaction by a parabola at its minimum shows that the
corresponding frequency is $\Omega/\omega =
\sqrt{2V_0/(\hbar\omega)}$.  Again, we find that the ground-state
wave function has a Gaussian shape in the direction of the
center-of-mass coordinate $(x_1+x_2)/\sqrt{2}$.  Figure~\ref{fig:afterfig5}
quantifies this statement. The circles show the relative error of the
ground-state energy (obtained from comparing the result in a model
space of $N=8$ oscillator shells with the result in $N=16$ oscillator
shells) as the strength $V_0$ of the interaction is varied. The squares
(diamonds) show to what extent the ground-state wave function
factorizes in a model space consisting of $N=8$ ($N=16$) oscillator
shells. The residual $1-s_1^2$ is clearly very small, and it decreases
with increasing size of the model space.  The triangles show the
ground-state expectation value of the center-of-mass
Hamiltonian~(\ref{Hosc}), normalized by the corresponding spacing
$\hbar\tilde{\omega}$. Thus, the center-of-mass wave function
factorizes, and it is to a high accuracy a Gaussian.  The result
obtained for this simple Hamiltonian lends further support to the
results obtained in the coupled-cluster approach.  It thus seems that
one might obtain a very accurate factorization in sufficiently large
model spaces even without employing an $N\hbar\omega$ space. Again,
this factorization is not surprising since the model space becomes
more and more complete as it increases in size.  The emergence of a
Gaussian center-of-mass wave function, however, is remarkable and not
yet understood.

The center-of-mass problem in coupled-cluster calculations has also
recently been addressed by Roth and coworkers~\cite{RGP}.  These
authors consider a Hamiltonian of the form $H=H_{\rm in}+\beta H_{\rm
  cm}(\omega)$, that is, the center-of-mass Hamiltonian~(\ref{Hosc})
with a frequency fixed to that of the employed harmonic oscillator
basis is added to the intrinsic Hamiltonian. This method, pioneered by
Gloeckner and Lawson~\cite{Lawson}, aims at shifting spurious states
up in the spectrum. Roth {\it et al.} found that the coupled-cluster
wave function for such a Hamiltonian does not factorize into a
Gaussian with frequency $\omega$ for center-of-mass wave function and
an intrinsic wave function.  Two comments are in order. First, we
believe that this approach misses the point as it requires perhaps
more than the coupled-cluster approximation needs to deliver.  For
atomic nuclei, there is no center-of-mass interaction, and the
requirement that the wave function factorizes in presence of a
center-of-mass interaction is more than one needs. Some time ago,
McGrory and Wildenthal also pointed out that the method by Gloeckner
and Lawson is not fully appropriate in truncated model spaces that are not
complete $N\hbar\omega$ spaces~\cite{McGrory}.  For nuclei, one only
needs that the wave function of the intrinsic Hamiltonian, when
computed in a Hilbert space of $A$ coordinates, exhibits a
factorization. Second, we caution that the conclusions by Roth and
coworkers are based on results that are not fully converged with
respect to the size of the model space (see Sect.~\ref{sec:ucom}
below).

The results of this section suggest the following procedure when
dealing with model spaces that are not complete $N\hbar\omega$ spaces:
(i) Compute the spectrum of the intrinsic Hamiltonian $H_{\rm in}$ in
as large a model space as conveniently possible. (ii) Check {\em a posteriori}
whether the resulting center-of-mass wave function is a Gaussian by
computing the expectation value $E{\rm cm}$, and determine the
corresponding frequency $\tilde\omega$. (iii) The ratio $E_{\rm
  cm}/(\hbar\tilde\omega)$ serves as a conservative estimate for the
quality of the factorization. (Strictly speaking, this ratio measures
to what extent the center-of-mass wave function deviates from a
Gaussian with frequency $\tilde\omega$.)

\section{Coupled-cluster results for medium-mass nuclei with chiral
  interactions}
\label{sec:medmass}

In this section, we compute binding energies and radii of several
doubly magic nuclei within the CCSD and the $\Lambda$-CCSD(T)
approximation, and employ chiral NN interactions.

Our single-particle wave functions are eigenfunctions of the harmonic
oscillator and characterized by the frequency $\omega$. Our model
space consists of spherical oscillator wave functions with radial
quantum number $n$ and angular momentum $l$, and we include
single-particle states with $2n+l\le N$ and $l\le 10$.  Fully
converged energies have to be independent of the parameters $N$ and
$\omega$ of our single-particle basis. In practice, we cannot go to
infinitely large spaces, but this is also not necessary for the
description of low-energy properties of finite nuclei. Our basis needs
to be complete in the following sense. It must be sufficiently
extended in momentum space to resolve the cutoff $\Lambda_\chi$ of the
employed interaction, and its extension in position space must be such
that a nucleus of radius $R$ literally fits into the basis. Let us
estimate the parameters required for a model space of oscillator
functions.

An oscillator basis consisting of $N$ shells at a frequency $\omega$ 
resolves the high momentum cutoff $\lambda$ if the
inequality 
\be
\label{uv}
N\hbar\omega \gtrsim {\hbar^2\lambda^2\over m}
\ee
is fulfilled. Likewise, the basis has to be sufficiently extended in 
position space to describe 
a nucleus with a radius $R$. For this, the inequality 
\be
\label{ir}
\hbar\omega \lesssim N{\hbar^2\over mR^2} 
\ee
needs to be fulfilled. In other words, an oscillator basis with
oscillator length $l_{\rm ho}=\sqrt{\hbar/(m\omega)}$ exhibits the
infrared cutoff $l_{\rm ho}^{-1}/\sqrt{N}$ and the ultraviolet cutoff
$l_{\rm ho}^{-1}\sqrt{N}$. Thus, converged results require model space
parameters $(N,\omega)$ that fulfill the inequalities (\ref{uv}) and
(\ref{ir}), and the results will then be insensitive to the specific
values of the parameters. Note that the simultaneous fulfillment of
Eq.~(\ref{ir}) and Eq.~(\ref{uv}) requires $N\gtrsim \lambda R$. (Our
calculations presented in this paper show that these approximate
relations are reasonable estimates.) Note also that
\be 
\label{hbaromega}
\hbar\omega\approx{\hbar^2\lambda\over mR}, 
\ee 
for a minimum model space with $N\approx\lambda R$.  Thus, the
well-known estimate $\hbar\omega\approx 42/A^{1/3}$~MeV is only valid
for small cutoffs $\lambda\approx k_F$ that are close to the Fermi
momentum $k_F$. These considerations show that much is to be gained
from low-momentum interactions, and -- conversely -- that ``bare''
chiral interactions with $\lambda=\Lambda_\chi$ require very large
model spaces.

Let us consider the oscillator basis as an example. A model space of
$N$ oscillator states contains about $\Omega\approx N^3/3$
single-particle states.  We obtain converged results for $^{16}$O
and the interactions from chiral EFT in
$N\approx 15$ shells. In this model space there are
$\Omega!/((\Omega-Z)!Z!)\approx 10^{20}$ Slater determinants for the
protons alone. Thus, the resulting model space for protons and
neutrons is far out of reach from diagonalization methods.

We employ the chiral NN interaction at order N$^3$LO, i.e., the NN
interaction is included up to the order $(Q/\Lambda_\chi)^3$, while 3NF
at this order and interactions at higher order are neglected.  We
perform a spherical Hartree-Fock calculation and transform the
Hamiltonian to the Hartree-Fock basis. The CCSD equations are then
solved in the spherical Hartree-Fock basis, and the Hartree-Fock state
is taken as the reference state for the coupled-cluster method. Note
that the chiral N$^3$LO interaction is quite ``hard''. Within the
Hartree-Fock approximation one does not even obtain bound nuclei.
Nevertheless, the solution of the non-linear coupled-cluster equations
yields rather well-bound nuclei.

\begin{figure}[t]
\includegraphics[width=0.45\textwidth,clip=]{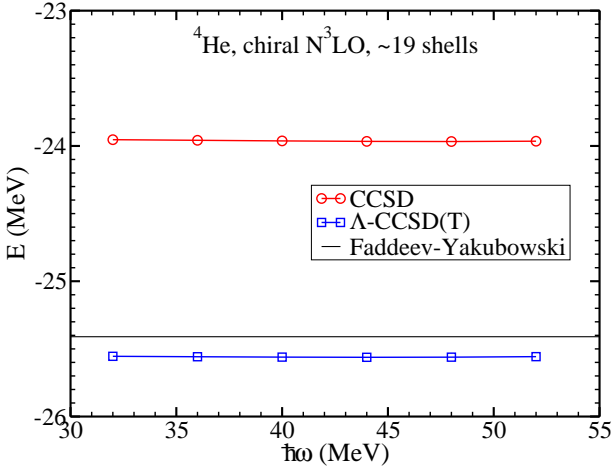}
\caption{(Color online) CCSD and $\Lambda$-CCSD(T) ground state energy
  of $^4$He using a chiral
  NN interaction at order N$^3$LO in 19 major oscillator shells as a function of 
  the oscillator spacing $\hbar\omega$, compared to virtually exact results from 
  Faddeev-Yakubowsky calculations.}
\label{fig:fig6}
\end{figure}

Fig.~\ref{fig:fig6} shows the ground-state energy of $^4$He in a
model space of $N=18$ (19 major oscillator shells). In such a large
model space, the results are virtually independent of the frequency
$\omega$ over a wide range.  The CCSD result for $^4$He deviates from
the virtually exact result from the solution of the Faddeev-Yakubowsky
(FY) equations by about 6\%. This difference is due to the omission of
three- and four-particle clusters. The more accurate $\Lambda$-CCSD(T)
method includes three-body clusters approximately and overshoots the
(FY) result by about 1.5\%. This overbinding is due to the
non-variational character of the coupled-cluster method and is not
really a concern due to the accuracy of the method.  The experimental
ground-state energy of $^4$He is $E=-28.3$MeV, and the additional
binding must be attributed to the 3NFs at order N$^3$LO; other
high-order terms in the chiral Langrangian play only a very small role
in $^4$He.

\begin{figure}[t]
\includegraphics[width=0.45\textwidth,clip=]{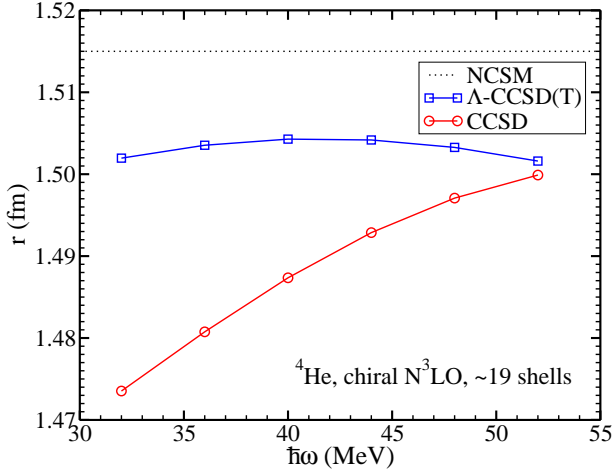}
\caption{(Color online) $^4$He RMS radii calculated using the Hellmann-Feynman
theorem within  the CCSD and the $\Lambda$-CCSD(T) approximations using a chiral
NN interaction at order N$^3$LO with cutoff 500~MeV$c^{-1}$. 
Calculations employed 
19 major oscillator shells, and the results are 
plotted as a function of the oscillator spacing $\hbar\omega$. }
\label{fig:fig7}
\end{figure}

We also computed the radius of the alpha particle within the
Hellmann-Feynman approach.  The results shown in Fig.~\ref{fig:fig7}
exhibit a very weak dependence on the oscillator frequency,
particularly within the $\Lambda$-CCSD(T) approximation. Within this
approach, the radius deviates by about 0.01~fm from the NCSM
results~\cite{NavRad07}, and about 0.015~fm from the virtually exact
result from the hyperspherical harmonics method employed by the Pisa
group~\cite{Pisa}.  We believe that the results presented for the
alpha particle manifest the high degree of accuracy that the coupled-cluster 
approximation exhibits. Note that the alpha particle also
exhibits the approximate factorization into a Gaussian center-of-mass
wave function and an intrinsic wave function. Here, the frequency of
the Gaussian is $ \hbar\tilde\omega\approx 19$~MeV while the
expectation value of the center-of-mass Hamiltonian is much smaller
$E_{\rm cm} \approx 0.3$~MeV.

\begin{figure}[h]
\includegraphics[width=0.45\textwidth,clip=]{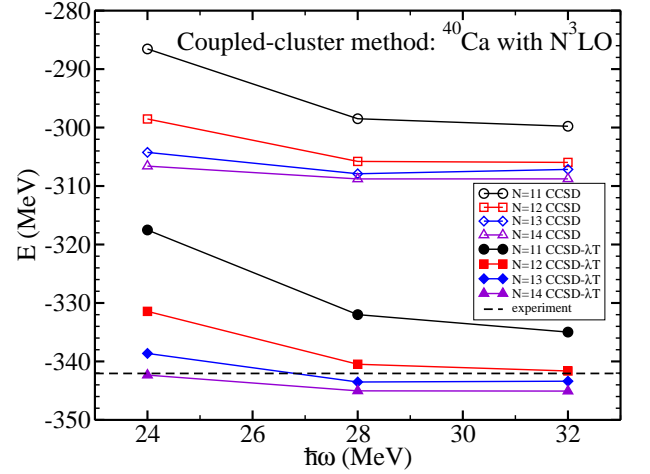}
\caption{(Color online) $\Lambda$-CCSD(T) and CCSD results for $^{40}$Ca from the chiral
NN interaction at order N$^3$LO as a function of the
oscillator spacing $\hbar\omega$ and the size of the model space.}
\label{fig:fig8}
\end{figure}

We next turn to the calcium isotopes $^{40}$Ca and $^{48}$Ca.
Figure~\ref{fig:fig8} shows the CCSD and $\Lambda$-CCSD(T) ground state
energy for $^{40}$Ca as a function of the oscillator frequency $\hbar\omega$ 
and the size of the model space $N$ (here the number of major shells is $N+1$). 
For $\hbar\omega = 32$~MeV and $N+1=15$ shells the 
$\Lambda$-CCSD(T) ground state energy is $-345.074$~MeV. We also performed a 
calculation in $N+1=19$ shells at $\hbar\omega = 32$~MeV, 
yielding a $\Lambda$-CCSD(T) ground state energy 
of $-345.781$~MeV. This shows that our results are converged within 1-2 MeV 
with respect to the size of the model space.  
Within this approximation, one overbinds $^{40}$Ca by about 3~MeV.  Compared to the previously
published CCSD results, the triples corrections add more than 30~MeV
of binding.

\begin{figure}[h]
\includegraphics[width=0.45\textwidth,clip=]{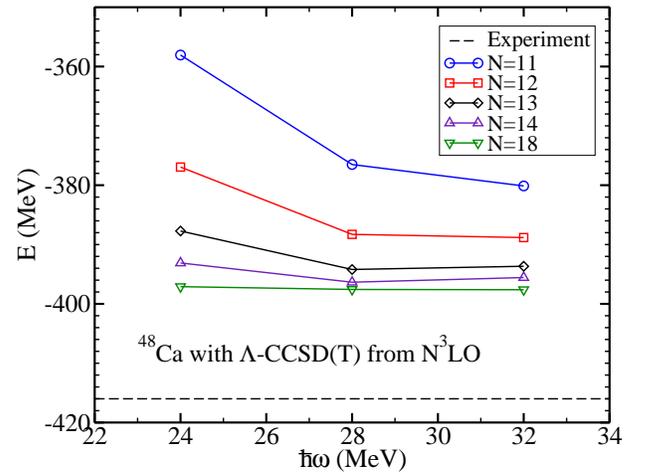}
\caption{(Color online) $\Lambda$-CCSD(T) and CCSD results for $^{48}$Ca from the chiral
NN interaction at order N$^3$LO as a function of the
oscillator spacing $\hbar\omega$ and the size of the model space.}
\label{fig:fig9}
\end{figure}

Figure~\ref{fig:fig9} shows the $\Lambda$-CCSD(T) results for $^{48}$Ca.
The results are converged to within about 2~MeV in a model space of
$N+1=19$ shells, and only exhibit a weak model-space dependence in the
largest model space. The approximation with triples yields about an
additional 40~MeV of binding when compared to the CCSD results. Let us
assume that the difference between our calculations and the experiment
is mainly due to the omitted 3NFs.  This suggests that the three-nucleon forces (3NF) will
exhibit an interesting isospin dependence. For $^{40}$Ca, only small
(repulsive) contributions from the 3NF are expected, while about
20~MeV of attraction is needed for $^{48}$Ca. A similar picture
arises when comparing $^{16}$O with $^{22}$O. For the former, about
0.41~MeV per nucleon in binding are missing, while the latter lacks
about 0.82~MeV per nucleon in binding~\cite{Hag09_ox}.

At this point, it is interesting, to compare the energies from the
Hartree-Fock calculation, the CCSD calculation, and the
$\Lambda$-CCSD(T) approximation. This comparison will permit us to
estimate the precision of the coupled-cluster method.
Table~\ref{tab1} shows that CCSD accounts for about 90\% of the
correlation energy, while the $\Lambda $-CCSD(T) approximation yields
about 10\%. This hierarchy has been observed in quantum chemistry as well, 
see for example Ref.~\cite{Bar07}. For
the alpha particle we know that four-particle clusters account again
for about 10\% of the triples correction.  Thus, it makes no sense to
artificially increase the precision of our results by turning to even
larger model spaces. The error estimates due to the finite model space
are of the same order as the estimates due to omitted four-body
clusters. Note that the results in Table~\ref{tab1} are also
consistent with the size extensivity of the employed coupled-cluster
methods, as the deviation $\Delta E$ from data is approximately linear
in mass number, i.e., $\Delta E/A$ is approximately constant over a
considerable range. 

\begin{table}[h]
  \begin{tabular}{|c|r|r|r|r|}
    \multicolumn{1}{c}{} & \multicolumn{2}{c}{ CCSD }&
    \multicolumn{2}{c}{ $\Lambda$-CCSD(T) } \\
    \hline
    Nucleus   & $E/A$   & $\Delta E / A$ &
    $E/A$   & $\Delta E / A$ \\ \hline
    $^{16}$O  &  -6.72  & 1.25 & -7.56  &  0.41  \\\hline
    $^{40}$Ca &  -7.72  & 0.84 & -8.63  & -0.08  \\\hline
    $^{48}$Ca &  -7.40  & 1.27 & -8.26  &  0.40 \\\hline
  \end{tabular}
  \caption{Ground-state energies per nucleon $E/A$ and deviation $\Delta E/A$ from experiment for doubly magic nuclei within the CCSD and $\Lambda$-CCSD(T) approximations. All energies are in units of MeV.}
  \label{tab1}
\end{table}

It is also very interesting to compare our results with the
results~\cite{Fujii2009} by Fujii {\it et al.} obtained within the
Unitary Model Operator Approach~\cite{UMOA}. For $^{16}$O, Fujii~{\it
  et al.}  report binding energies of 6.62~MeV per nucleon and
7.47~MeV per nucleon employing two-body clusters and three-body
clusters, respectively. The close agreement between our results
presented in Table~\ref{tab1} and those of Ref.~\cite{Fujii2009}
demonstrate that different {\it ab initio} methods are setting
reliable benchmarks for increasingly heavier nuclei.

\section{Shell evolution with chiral NN interactions}
\label{sec:evolve}
In this section we investigate the evolution of the low-lying positive
parity states ${1/2}^+$, ${3/2}^+$, and ${5/2}^+$ in the oxygen and
fluorine isotope chains using the chiral nucleon-nucleon interaction
at N$^3$LO.  Our calculations do not include three-body forces, but in
order to probe the effects of omitted short-ranged many-body forces we
renormalize the ``bare'' chiral interaction using the similarity
renormalization group method and study the resolution scale
dependence on the calculated energies.  The low-lying
states $J^\pi={1/2}^+$, $J^\pi={3/2}^+$, and $J^\pi={5/2}^+$ in oxygen and
fluorine are usually interpreted as the $s_{1/2}$, $d_{3/2}$, and
$d_{5/2}$ proton and neutron single-particle states in the closed-shell 
oxygen isotopes $^{16}$O, $^{22}$O, $^{24}$O, and $^{28}$O. Here,
we define the single-particle energy of the state $J^\pi$ as the
difference in binding energy between the $A\pm 1$ nucleus and the closed-shell
nucleus $A$, i.e., 
\be
\label{esp}
E_{\mathrm sp} (J^\pi)= E^{A\pm 1}(J^\pi) -E_0^A \ .
\ee

The evolution of nuclear shell structure for neutron-rich isotopes is
of great theoretical and experimental interest, see for example the recent
review~\cite{Sorlin}.  Some of the traditional magic numbers of the
nuclear shell model might fade away as one moves away from the valley
of beta stability, while new magic numbers emerge in neutron-rich
nuclei. The evolution of shell structure in the isotopes of oxygen has
recently received considerable
experimental~\cite{thirolf2000,stanoiu2004,kanungo2009,hoffman2009}
and theoretical
attention~\cite{Luo,Michel_oxygen,Volya,Hag09_ox,taka09,Koshiroh09}.
Considerable shell gaps have been observed in $^{22}$O
\cite{thirolf2000} and $^{24}$O \cite{stanoiu2004,kanungo2009},
leading to the interpretation of new magic numbers at ($Z=8,N=14$) and
($Z=8,N=16$), respectively.  Hoffman {\it et al.}~\cite{hoffman2009}
found that $^{25}$O is a resonance and unstable towards one-neutron
emission.  The instability of $^{25}$O is closely related to the
location of the $d_{3/2}$ single-particle shell, and it is clear that
the evolution of the $d_{3/2}$ shell will decide whether $^{24}$O or
$^{28}$O is the most neutron-rich stable isotope of oxygen.

The underlying microscopic mechanisms for the evolution of nuclear
shell structure is not yet entirely understood.  One might expect that
both three-nucleon forces and coupling with the scattering continuum
will play significant roles in predicting the limits of nuclear
stability and the shell evolution towards the drip line.
Zuker~\cite{zuker03} suggested that three-nucleon forces modify the monopole terms of
microscopically derived shell-model interactions.  Indeed, Otsuka {\it
  et al.}~\cite{taka09} found within the $sd$ shell model that 3NFs
will add repulsion between the $d_{5/2}$ and $d_{3/2}$ orbitals, thus
making $^{24}$O the heaviest bound isotope of oxygen.  Coupled-cluster
calculations of the isotopes $^{16,22,24,28}$O reveal that
three-nucleon forces (or more complicated many-body forces) play a 
non-negligible role in the determination of the drip-line for oxygen isotopes.

In this section, we study the evolution of shell structure in the
isotopes of oxygen and fluorine based on chiral NN
interactions~\cite{N3LO}. We will omit continuum effects and 3NFs
from our study. Recall that chiral 3NFs consist of a long-ranged
two-pion exchange, a mid-range one-pion exchange, and a contact
term~\cite{Nogga3NF}. Variation of the cutoff will induce short-range
3NFs that are in their structure identical to the chiral
3NF~\cite{NBS,Juergenson}. Thus, we will be able to probe the effect
of short-ranged 3NFs {\it a posteriori} by cutoff variation.  Continuum
effects are, of course, expected to be important for nuclei close to
the drip line~\cite{michel07,quaglioni08,hagen10}. However, it will
turn out that most of the computed single-particle states are fairly
well bound (likely due to the omission of 3NFs). In this situation,
the continuum plays a smaller role.
 
\begin{figure}[h]
\includegraphics[width=0.45\textwidth,clip=]{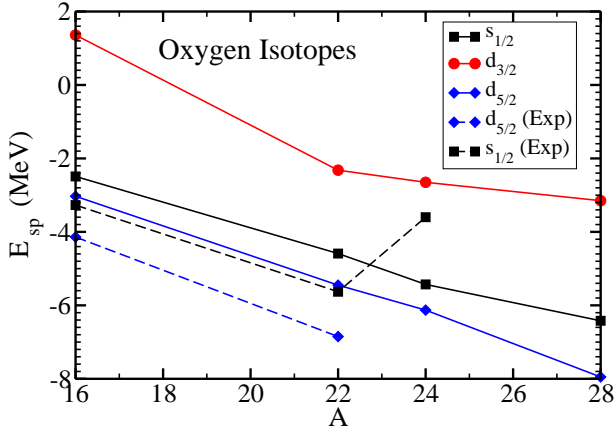}
\caption{(Color online) Single-particle energies (Eq.~(\ref{esp})) of the $s_{1/2}, d_{3/2}$, and the $d_{5/2}$ 
  single-particle states in the oxygen isotopes using the 
  ``bare'' chiral NN interaction. The calculations employ 
  $N = 14$ shells for fixed $\hbar\omega= 28$ MeV. The $d_{5/2}$ and
  $s_{1/2}$ experimental single-particle energies are shown as dashed lines.}
\label{fig:fig_1}
\end{figure}

\begin{figure}[h]
\includegraphics[width=0.45\textwidth,clip=]{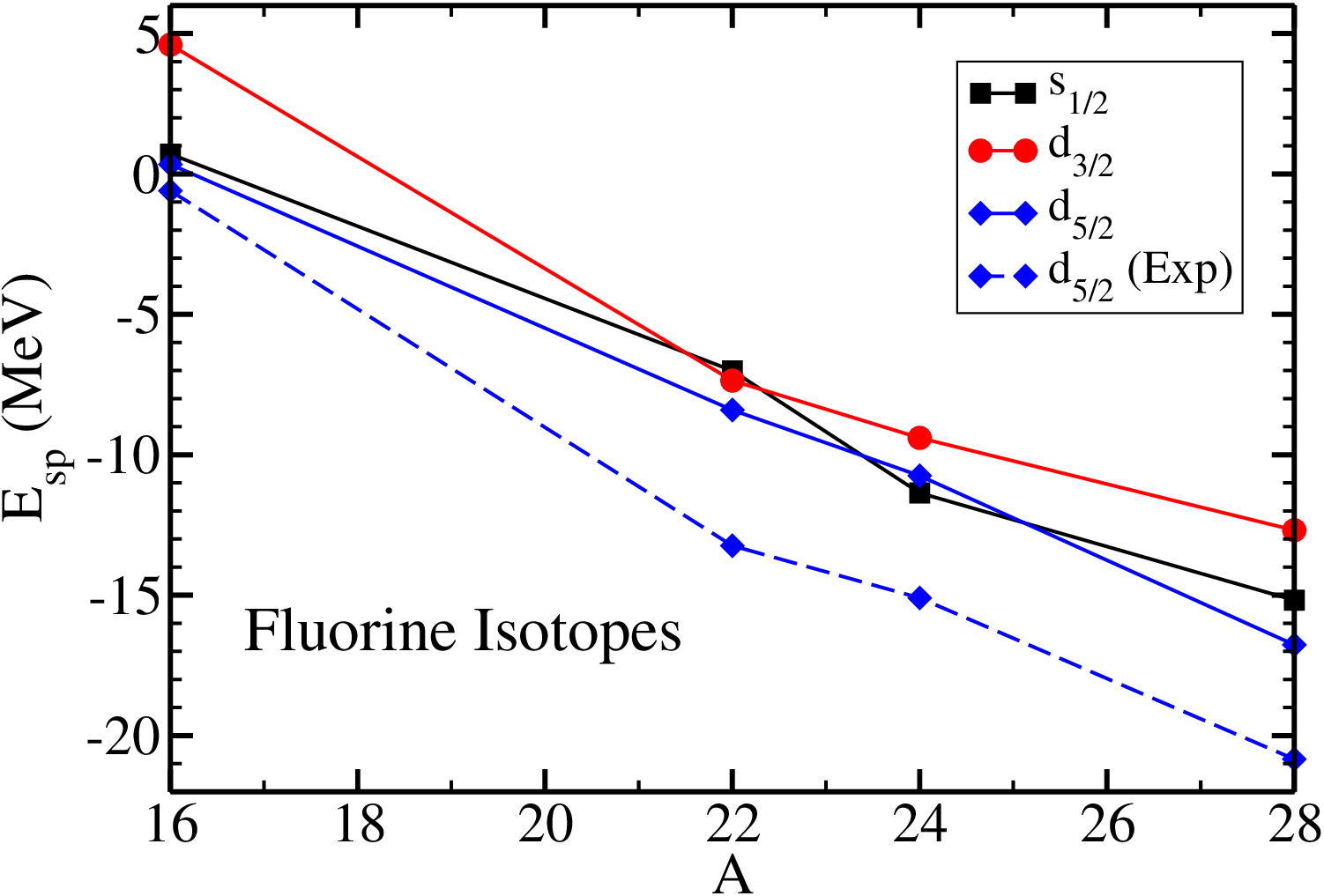}
\caption{(Color online) Single-particle energies (Eq.~(\ref{esp})) of the $s_{1/2}, d_{3/2}$, and the $d_{5/2}$ 
  single-particle states in the fluorine isotopes using the 
  ``bare'' chiral NN interaction. The calculations employ 
  $N = 14$ shells for fixed $\hbar\omega= 28$ MeV. The experimental $d_{5/2}$ single-particle
  energiy is shown as a dashed line.}
\label{fig:fig_2}
\end{figure}

Figures~\ref{fig:fig_1} and ~\ref{fig:fig_2} show the effective
single-particle energies for the neutron and proton $s_{1/2}$,
$d_{3/2}$, and $d_{5/2}$ states in $^{16}$O, $^{22}$O, $^{24}$O, and
$^{28}$O.  The calculations employed a model space consisting of
$N+1=15$ oscillator shells at a fixed oscillator frequency
$\hbar\omega = 28$~MeV.  The proton $s_{1/2}, d_{3/2}$, and the
$d_{5/2}$ single-particle states were computed within the PA-EOM-CCSD
approach. The neutron single-particle states resulted from either the
PR-EOM-CCSD or the PA-EOM-CCSD method (See subsection \ref{subsec:eomccsd} for more details), 
depending on whether the
considered single-particle state is a hole or a particle state of
an isotope of oxygen with a closed subshell. 

Several interesting features can be extracted from the results.  The
computed neutron single-particle energies (shown in
Fig.~\ref{fig:fig_1}) do not reproduce the experimentally observed
shell gaps in $^{22}$O and $^{24}$O.  Furthermore, the chiral NN
interactions incorrectly yield a bound $^{25}$O~\cite{hoffman2009}.  This
behavior has also been seen in shell-model calculations of the oxygen
isotopes using microscopically derived shell-model interactions (see
for example Ref.~\cite{taka09}).  For the fluorine isotopes (shown in
Fig.~\ref{fig:fig_2}), the $d_{5/2}$ state follows the experimental
trend but lacks binding. In $^{25}$F our calculations yield an
inversion of the $s_{1/2}$ and $d_{5/2}$ states, giving a ${1/2}^+$
state as the ground state of $^{25}$F.  In order to gauge the effects
of omitted short-range 3NFs, we vary the resolution scale $\lambda$ by
a similarity renormalization group (SRG) transformation~\cite{SRG}.

\begin{figure}[h]
\includegraphics[width=0.45\textwidth,clip=]{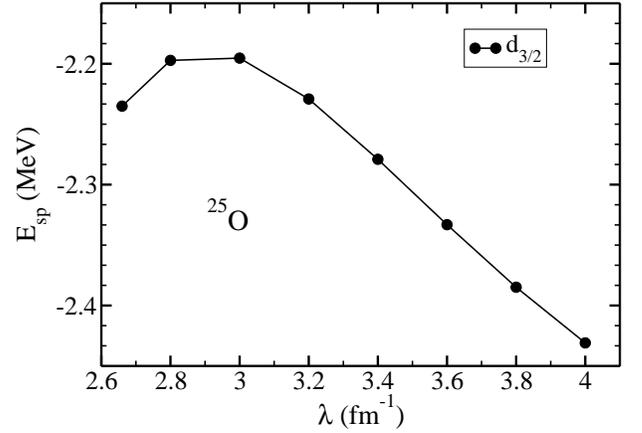}
\caption{(Color online) 
  Single-particle Energy (Eq.~(\ref{esp})) of the neutron $d_{3/2}$ state in $^{24}$O 
  as a function of the resolution scale $\lambda$ of the  SRG evolved chiral NN
  interaction. The calculations employ
  $N = 14$ shells at fixed $\hbar\omega= 26$ MeV.} 
\label{fig:fig_3}
\end{figure}

Figure~\ref{fig:fig_3} shows the dependence of the
neutron $d_{3/2}$ single-particle state in $^{24}$O as a function of
the SRG cutoff $\lambda$. In these calculations we used
$N+1=15$ oscillator shells with a fixed oscillator frequency of
$\hbar\omega = 26$MeV.

\begin{figure}[h]
\includegraphics[width=0.45\textwidth,clip=]{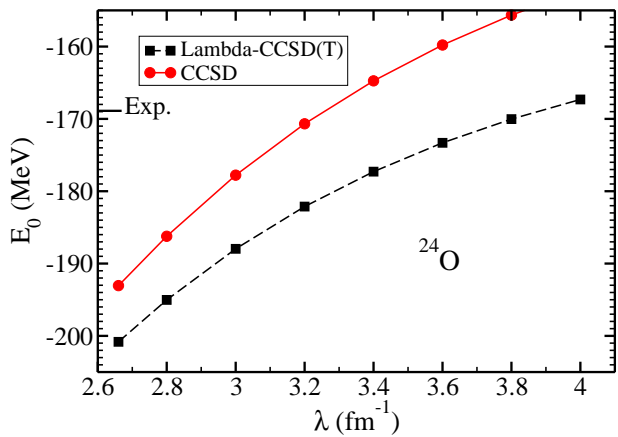}
\caption{(Color online) Ground-state energy of  $^{24}$O within the CCSD (circles) and the $\Lambda$-CCSD(T) 
  approximation (squares) as a function of the resolution scale $\lambda$ of the  SRG evolved chiral NN
  interaction. The calculations employ
  in $N +1= 15$ shells at fixed $\hbar\omega= 26$~MeV.}
\label{fig:fig_5}
\end{figure}

\begin{figure}[h]
\includegraphics[width=0.45\textwidth,clip=]{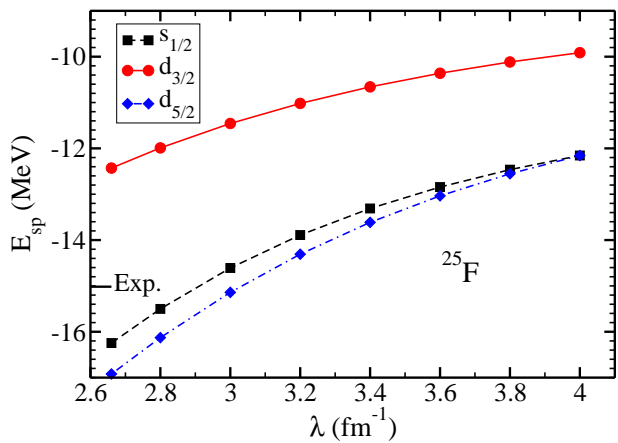}
\caption{(Color online)
  Single-particle energies (Eq.~(\ref{esp})) of the proton $s_{1/2}, d_{3/2}$, and $d_{5/2}$ states 
  in $^{24}$O as a function of the resolution scale $\lambda$ of the  SRG evolved chiral NN
  interaction. The calculations employ
  $N+1 = 15$ shells at fixed $\hbar\omega= 26$~MeV.} 
\label{fig:fig_4}
\end{figure}

The $d_{3/2}$ single-particle state in $^{24}$O depends weakly on the
cutoff and stays bound for all choices of the cutoff $\lambda$. This
suggests that long-ranged 3NFs are needed to yield an unbound state
and to place the drip line at $^{24}$O.  Furthermore, the
effect of the scattering continuum is expected to add some additional
binding to this state (see Ref.~\cite{hagen10}). Thus, a realistic
description of $^{25}$O will result from a fine interplay between the
scattering continuum and 3NFs. Figure~\ref{fig:fig_5} shows the cutoff
dependence of the CCSD and $\Lambda$-CCSD(T) ground-state energies for
$^{24}$O (using the same model space and oscillator frequency as in
Fig.~\ref{fig:fig_3}). Note that there is no cutoff $\lambda$ that
simultaneously reproduces the experimental binding energy of $^{24}$O
and the resonance energy of $^{25}$O ground state.
Figure~\ref{fig:fig_4} shows that proton $s_{1/2}, d_{3/2}$, and
$d_{5/2}$ single-particle states in $^{24}$O exhibit a considerable
dependence on the cutoff $\lambda$.  No single cutoff reproduces
simultaneously the ground-state binding energies of $^{24}$O, $^{25}$O,
and $^{25}$F, again pointing to the importance of the omitted 3NFs and
to subtleties in its isospin dependence.

In conclusion, chiral NN interactions alone do not reproduce the
evolution of single-particle energies within the $sd$ shell. Close to
$^{16}$O, the single-particle states are in semi-quantitative
agreement with data, but missing 3NFs become increasingly more
important for neutron-rich isotopes of oxygen. It is expected, that a
realistic description of single-particle states and the evolution of
shell structure will require a theory which allows for a consistent
and systematic inclusion of many-body correlations, 3NFs, and coupling
to the scattering continuum. The beauty of coupled-cluster theory is
that it allows for inclusion of all these ingredients in a simple and
transparent way, and we aim to investigate and predict properties of
nuclei from the valley of stability to the very limits of matter
taking all these ingredients into account in the near future.

\section{Resolution-scale dependence for $^{40}$Ca and power counting}
\label{sec:srg}
In this section we study the ground-state properties of $^{40}$Ca
using renormalized nucleon-nucleon interactions derived with the SRG
method~\cite{SRG}. The SRG method drives the Hamiltonian to a band
diagonal form, and therefore decouples low-momentum degrees of freedom
from high-momentum degrees of freedom. The cutoff parameter $\lambda $
determines the decoupling and sets the resolution scale or energy
scale with which we can probe the structure of a particular nucleus.
However, this procedure induces three-body forces and forces of
higher rank. At the two-body level, these interactions typically
overbind medium-mass nuclei considerably, and show a strong
dependence on the cutoff $\lambda$. The dependence on the cutoff
$\lambda$ on calculated observables gives an indication of the missing
physics and on the rol e of many-body forces not included in the
calculation; only the sum of all forces induced by the renormalization
is truly independent of the scale of resolution or the cutoff.

For the ground-state calculations of $^{40}$Ca, we use SRG
interactions at the resolution scales $\lambda=2.5, 2.2, 1.9$
fm$^{-1}$, and compute the binding energy within the CCSD
approximation. We solve the CCSD equations in a spherical harmonic
oscillator basis. The SRG interactions are soft and the corrections
due to triples clusters are found to be rather small.  For $^{40}$Ca
and a low-momentum interaction $\vlowk$ with cutoff
$\lambda=1.9$fm$^{-1}$, for instance, the CCSD binding energy is
$E/A=12.28$ MeV and the CCSD(T) approximations yields an additional
0.29 MeV per nucleon~\cite{Hag07_benchmark}. Therefore, we limit the
computations for the SRG interactions to the CCSD approximation.  Our
focus is on the saturation and convergence properties of the SRG
interactions in medium-mass nuclei and not on precision results.

\begin{figure}[h]
\includegraphics[width=0.45\textwidth,clip=]{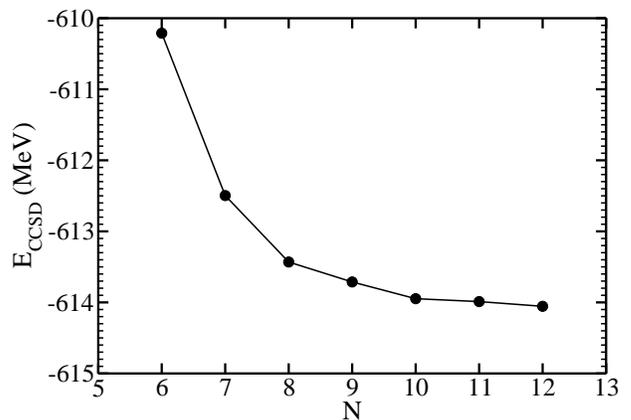}
\caption{(Color online) CCSD ground-state calculation of $^{40}$Ca
 for increasing number of oscillator shells, $N=2n+l$, at fixed $\hbar\omega =
 26$MeV, using SRG evolved chiral NN
  interaction at resolution scale $\lambda=1.9$fm$^{-1}$.}
\label{fig:fig10}
\end{figure}

\begin{figure}[h]
\includegraphics[width=0.45\textwidth,clip=]{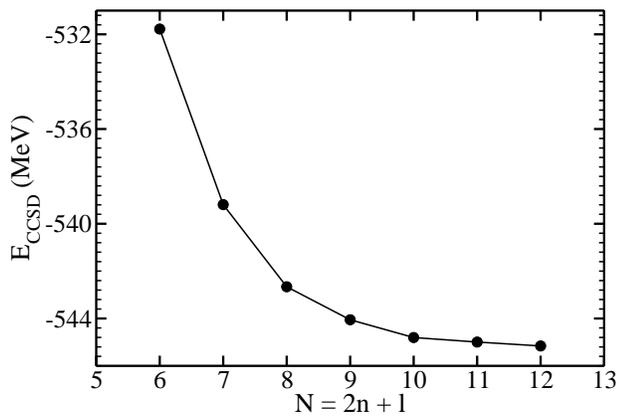}
\caption{(Color online) Same caption as in Fig. \ref{fig:fig10} except that resolution
 scale is $\lambda=2.2$fm$^{-1}$.}
\label{fig:fig11}
\end{figure}

\begin{figure}[h]
\includegraphics[width=0.45\textwidth,clip=]{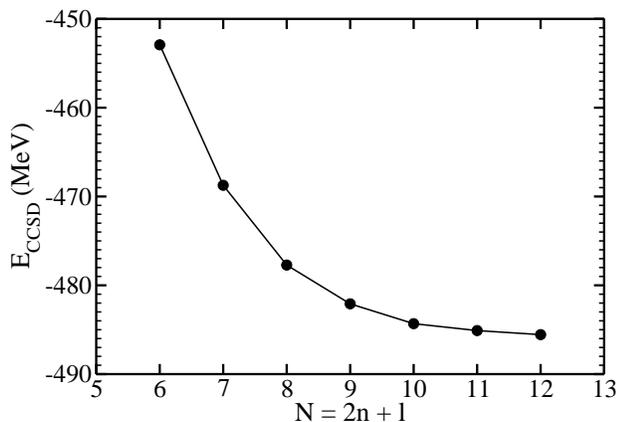}
\caption{(Color online) Same caption as in Fig. \ref{fig:fig10} 
  except that resolution
 scale is $\lambda=2.5$fm$^{-1}$.}
\label{fig:fig12}
\end{figure}

Figures \ref{fig:fig10}, \ref{fig:fig11}, and \ref{fig:fig12} show the
convergence of the ground-state energy of $^{40}$Ca using SRG evolved
interactions with resolution scales $\lambda=1.9, 2.2, 2.5$ fm$^{-1}$,
respectively. The ground-state energies computed for the various
cutoffs are well converged in model space sizes comprising up to 13
major oscillator shells. As the cutoff $\lambda$ is lowered, these SRG
interactions become increasingly soft, and we observe a faster
convergence with increasing basis size. The computed ground-state
energies show that the SRG NN interactions alone overbind considerably
with respect to experiment. The results also exhibit a strong
dependence on the resolution scale $\lambda$, and the overbinding with
respect to the experimental binding energy of $^{40}$Ca, which is
approximately $342$ MeV, increases with decreasing momentum scale.
This strong resolution dependence is, of course, due to our omission
of three-nucleon forces 3NFs (and more complicated many-body forces)
and higher-order terms~\cite{NBS,Juergenson}.  Likewise, with
increasingly softer NN interactions, the proper reproduction of
saturation properties must come from the repulsive character of the
3NF~\cite{taka09}.

Table~\ref{tab:tab2} shows the binding energy per particle $E/A$,
interaction energy per particle $V/A$, the momentum expectation value
$Q$, and the difference in binding energy per particle compared to
experiment $\Delta E/A$ for $^{40}$Ca using the different SRG
interactions.
\begin{table}[h]
\begin{tabular}{|c|r|r|r|r|c|}
  \hline
  $\lambda$            & $E/A$   & $V/A$  & $Q$  &$\Delta E/A$&$\left|{\Delta E\over V}\right|/\left(Q\over\lambda\right)^3$\\\hline
  1.9                  & -15.35  & -47.59 & 1.25 &  -6.80      & 0.50\\
  2.2                  & -13.63  & -44.84 & 1.23 &  -5.08      & 0.62\\
  2.5                  & -12.14  & -42.39 & 1.21 &  -3.59      & 0.77\\\hline
\end{tabular}
\caption{CCSD results for $^{40}$Ca with SRG interactions evolved from
the chiral N$^3$LO nucleon-nucleon interaction to the momentum scale
$\lambda$. The binding energy per nucleon, and interaction energy per
nucleon are denoted as $E/A$ and $V/A$, respectively. $Q$ denotes the
expectation value of the momentum probed in this nucleus, while
$\Delta E$ denotes the difference to the experimental binding
energy. Energies are in units of MeV and momenta in units of
fm$^{-1}$.}
\label{tab:tab2}  
\end{table}
To examine the power counting, we compute the expectation value of the
interaction energy via the Hellmann-Feynman theorem, and also deduce the
average momentum $Q$ from the expectation value of the kinetic energy.
The results are given in Table~\ref{tab:tab2}.  The CCSD results are very
well converged for momentum scales $\lambda=1.9, 2.2, 2.5$~fm$^{-1}$.
As already pointed out, our results are far from complete. Due to the
restriction to NN forces, we are missing contributions of 3NFs at
order N$^3$LO, and NN, 3NF, and 4NF forces from higher order. The
missing contributions from clusters due to the CCSD approximation are
much smaller than the contributions related to neglected terms of the
interaction. Chiral effective field theory puts the contributions of
3NFs at order $(Q/\lambda)^3$. The SRG transformation shifts
high-momentum NN contributions to forces of higher rank but is not
expected to destroy the power counting. Are our results consistent
with this expectation?  To answer this question, we compute the ratio
$\left|{\Delta E /V}\right|/\left(\lambda/ Q\right))^3$ for the
$^{40}$Ca nucleus and present the result in Table~\ref{tab:tab2}. The
ratios are of natural size and suggest that the power counting is not
violated in medium-mass nuclei.

Let us discuss the implications of this finding. From a theoretical
point of view, it is reassuring and satisfying that the power-counting
arguments seem applicable in medium-mass nuclei. From a practical
point of view, this raises the question about the order necessary to
achieve a desired precision. Low-momentum and SRG versions of NN
interactions are quite popular since they facilitate considerably the
solution of the nuclear many-body problem. However, our results also
suggest that the momentum expansion in $Q/\lambda$ has to be driven
considerably beyond the present order $(Q/\lambda)^3$ in order to
obtain a small error estimate due to missing contributions from the
nuclear interactions. Recall that the 4NF enters at order N$^3$LO.
Excluding cancellation among 4NFs (which might perhaps be achieved
through some fine-tuning of low-energy constants), the technical
advantage of soft NN interactions might interaction wise be offset due to
the technical difficulties that come with the inclusion of forces of
increasingly higher rank. This potential problem has not been noticed
in light nuclei. In light nuclei, the contributions of 3NFs are
reduced at rather small cutoffs due to fortuitous cancellations that
are not expected from the power counting. Here, the lower cutoff
benefits the handling of the NN force {\it and} it reduces the
contributions of 3NFs. In light systems, there also seems to be only
very little need for forces of rank higher than three.
Table~\ref{tab:tab2} hints that this might be different in $^{40}$Ca.
Here, and in agreement with the power counting, the contributions of
3NFs increase with the decreasing momentum scale of the underlying
interaction.

The results reported in this section suggest that there are at least
two possible routes to {\it ab-initio} calculations in medium-mass
nuclei. First, one can employ a low resolution scale $\lambda$ which
facilitates the solution of the nuclear many-body problem. The price
tag to be paid might consist of a slower convergence with respect to
the order $k_F/\lambda$, i.e., the evolution of four-body forces might
be needed.  Second, one can employ a high cut-off scale $\Lambda_\chi$ and
work on the technical difficulties related to the solution of the
nuclear many-body problem. In this case, the full N$^3$LO interaction
including 3NF forces might already suffice to achieve accuracies of
the order of 0.3 MeV per nucleon, and fine-tuning of poorly
constrained coefficients of the 3NF might further increase the
accuracy. In following the second path, it might be attractive to
employ other techniques to tame the ``hard'' NN interaction. In the
next section, we therefore explore the convergence properties of the
$G$-matrix computed from the chiral NN interaction at order N$^3$LO.

\section{Convergence properties of the $G$-matrix}
\label{sec:g}

In this section we calculate CCSD ground states of $^4$He and $^{16}$O
using the in-medium $G$-matrix \cite{mhj95}.  The $G$-matrix is an
effective in-medium interaction that is computed starting from the
bare nucleon-nucleon (NN) interaction using a Green's function
approach with unperturbed propagators \cite{greens}. It depends
therefore on the starting energy $\overline{\omega}$. The
starting energy defines the energy of the incoming and outgoing
single-particle states, and is normally set equal to the sum of the
unperturbed energies of the interacting single-particle states.  The
$G$-matrix is defined in a model space $P$ by summing ladder diagrams
to infinite order, where the intermediate particle states are defined
in the $Q$-space. Each ladder diagram scatters two particles from the
$P$-space to the complement $Q$-space and back to the $P$-space.  The
number of interaction vertices in a ladder diagram gives the number of
times the particles rescatter within the $Q$-space. It is clear that
the $G$-matrix depends inherently on the model space and the interaction. 
The purpose of the $G-$matrix is to tame the hard core of
the ``bare'' interaction making it suitable for many-body perturbation
theory.  In this section we investigate how the $G$-matrix behaves in
\emph{ab-initio} coupled-cluster ground-state calculations of $^4$He
and $^{16}$O as the model space $P$ increases and the starting energy
is varied over a wide range.

\begin{figure}[h]
\includegraphics[width=0.45\textwidth,clip=]{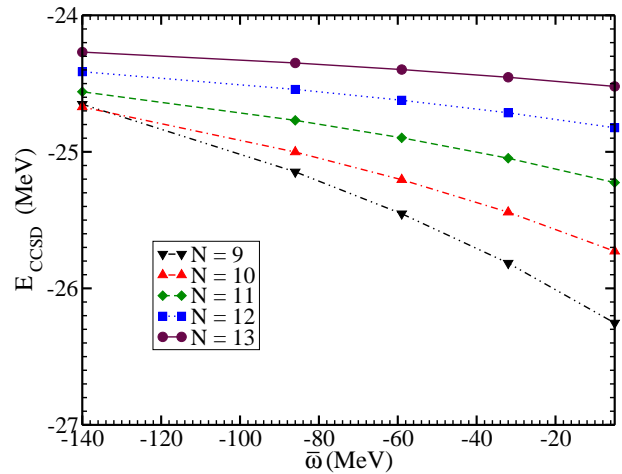}
\caption{(Color online) CCSD binding energy for $^{4}$He using a $G$-matrix
and starting energies $\overline\omega$ in the range from $-140$ MeV to $-5$ MeV. The oscillator spacing 
is held fixed at $\hbar\omega=20$~MeV.}
\label{fig:fig13}
\end{figure}

\begin{figure}[h]
\includegraphics[width=0.45\textwidth,clip=]{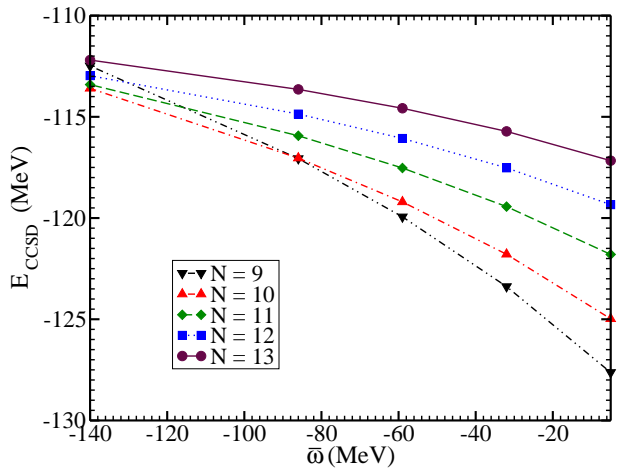}
\caption{(Color online) CCSD binding energy for $^{16}$O using a $G$-matrix
and starting energies $\overline\omega$ in the range from $-140$ MeV to $-5$ MeV. The oscillator spacing 
is held fixed at $\hbar\omega=20$~MeV.}
\label{fig:fig14}
\end{figure}

In constructing the $G$-matrix, we start from the ``bare'' N$^3$LO
interaction ($\Lambda_\chi = 500$~MeV$c^{-1}$) by Entem and Machleidt
\cite{machleidt02,N3LO} and a spherical harmonic oscillator basis. In
Figs.~\ref{fig:fig13} and ~\ref{fig:fig14}, we show the CCSD results
for $^4$He and $^{16}$O, respectively, starting from the calculated
$G$-matrix. In the calculations we varied the starting energy
$\overline\omega$ from -140 MeV to -5 MeV and studied the effect of
this variation on the CCSD ground state energy. We increased the model
space from 10 to 14 major oscillator shells while the oscillator
frequency was held fixed at $\hbar\omega = 20$~MeV. The figures show
that the dependence on the starting energy is reduced as we increase
the model space. In the largest model space considered here, the
starting-energy dependence is very mild. As $\overline\omega$ is
varied over a range of 135~MeV, the binding energy of $^4$He changes
only by about $0.25$ ~MeV and of $^{16}$O by about $5$~MeV.
Furthermore, we see that at a fixed starting energy, the binding
energy converges rather slowly with increasing size of the model
space. As the model space $P$ increases the complement space $Q$ in
which the particles scatter is reduced, and the $G$-matrix should
therefore converge to the underlying ``bare'' nucleon-nucleon
interaction in an infinite model space. This expectation is compatible
with our CCSD results for the ground-state energies of $^4$He (about
$-24$~MeV) and $^{16}$O (about $-108$~MeV) using the ``bare'' chiral
NN interaction (see, e.g., Figs.~\ref{fig:fig1} and ~\ref{fig:fig2}).
Figures~\ref{fig:fig13} and ~\ref{fig:fig14} show that the
ground-state energy of the $G$-matrix converges from below as the
model space is inceased in size, and that the result of the ``bare''
interaction is approached in very large model spaces.

In conclusion, we observe slow convergence for the CCSD ground-state
energy with increasing model space with the $G$-matrix defined from
the ``bare'' N$^3$LO interaction. The dependence on the starting
energy disappears in sufficiently large model spaces, and the
coupled-cluster results with the $G$-matrix converge towards
coupled-cluster results for the free nucleon-nucleon interaction. In
the case of the N$^3$LO interaction, it is clearly better to start from
the free interaction. It remains to be seen in the case of harder
interactions, like the Argonne V18 interaction, whether convergence is
better using the in-medium $G$-matrix as a starting point for
\emph{ab-initio} coupled-cluster calculations.

Let us briefly contrast the $G$ matrix to low-momentum NN interactions
{\`a} la $\vlowk$ or from the similarity renormalization
group. The latter two preserve phase shifts from NN scattering and
bound-state properties up to the chosen momentum cutoff. They soften
the NN interaction and -- when not augmented by the induced 3NF --
tend to considerably overbind heavier nuclei at lower cutoffs around
$\lambda\approx 2$~fm$^{-1}$ or so. In a finite model space, the
$G$-matrix always exhibits a dependence on the starting energy, it
converges slowly to the ``bare'' interaction, and it somewhat
overbinds the nuclear many-body system compared to the ``bare''
interactions that is used in its construction. The authors speculate
that this last point might seem attractive with a view on practical
applications, as ``bare'' nucleon-nucleon interaction models such as the Argonne
interaction, the CD-Bonn interaction and the chiral interaction~\cite{N3LO}
somewhat underbind nuclei. In other words, the slight overbinding of
the $G$-matrix often results into very reasonable energies for heavier
nuclei, and the $G$-matrix simply works very well in practical applications.

\section{Convergence properties of V$_{\mathrm{UCOM}}$}
\label{sec:ucom}

Let us also consider the saturation and convergence properties of the
V$_{\mathrm{UCOM}}$ nucleon-nucleon interaction~\cite{UCOM}.  The
interaction V$_{\mathrm{UCOM}}$ is obtained from a unitary correlation
operator method (UCOM) that softens the short-range repulsion of the
initial ``bare'' interaction by a similarity transformation with a
unitary operator, which is explicitly designed to remove the hard-core
and the short-ranged tensor components of phenomenological NN
interaction. In this section we consider the UCOM interaction obtained
from the ``bare'' Argonne V18 interaction.

The V$_{\mathrm{UCOM}}$ interaction was initially applied in
mean-field methods. These applications were very successful.  Very
reasonable binding energies and saturation properties could already be
obtained within the Hartree-Fock approximation. Only later did this
interaction see applications that included the inclusion of many-body
correlations within the RPA, many-body perturbation
theory~\cite{Roth06}, the coupled-cluster method~\cite{Roth09}, and
within the no-core shell model~\cite{Roth07}. (For a recent review, we
refer the reader to Ref.~\cite{UcomReview}.) Those later applications
suggest that the V$_{\mathrm{UCOM}}$ interaction is soft yet with
appealing saturation properties, and without the need for sizeable
3NFs. This situation seems surprising.  On the one hand, the
V$_{\mathrm{UCOM}}$ interaction is similar in its technical
construction to the SRG interactions~\cite{RothReinHer}. On the other
hand, the V$_{\mathrm{UCOM}}$ interaction seems to differ
significantly in its saturation property from the SRG nucleon-nucleon
interactions which tend to overbind nuclei at low momentum cutoffs.
We address this puzzle in what follows, and perform structure
calculations with the UCOM interaction in large model spaces and
for various oscillator frequencies.

\begin{figure}[h]
\includegraphics[width=0.45\textwidth,clip=]{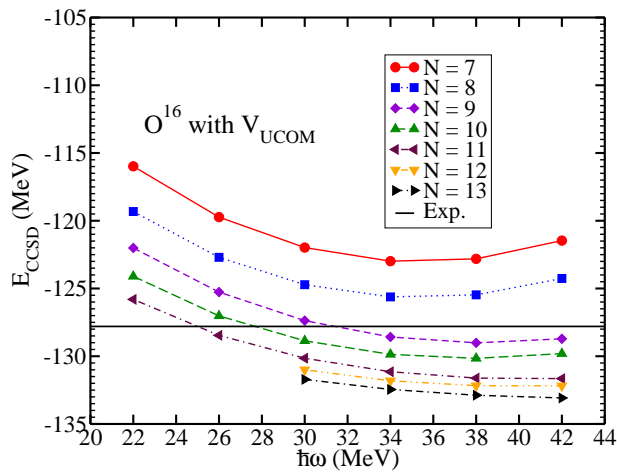} 
\caption{(Color online) CCSD ground state energy of $^{16}$O with
  $V_{{\rm UCOM}}$ as a function of the oscillator spacing $\hbar\omega$ and the size of
  the model space $N$. The experimental ground-state energy is given
  by the solid line.}
\label{fig:fig15}
\end{figure}

\begin{figure}[h]
\includegraphics[width=0.45\textwidth,clip=]{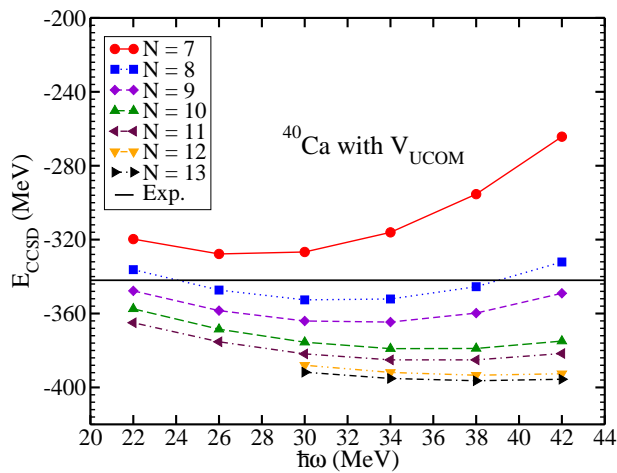}
\caption{(Color online) CCSD ground-state energy of $^{40}$Ca using
  $V_{{\rm UCOM}}$ as a function of the oscillator spacing
  $\hbar\omega$ and the size of the model space $N$. The experimental
  ground state energy is given by the solid line.}
\label{fig:fig16}
\end{figure}

We employ the V$_{\mathrm{UCOM}}$ interaction and compute the ground
states of $^{16}$O and $^{40}$Ca within the CCSD approximation. Our
calculations employ very large model spaces and a considerable range
of oscillator frequencies, and thereby differ from the previous
studies~\cite{Roth06,Roth07,Roth09}; see also
Ref.~\cite{Dean08_Comment}.  Figures~\ref{fig:fig15} and \ref{fig:fig16}
show the ground-state energy of $^{16}$O and $^{40}$Ca, respectively,
as a function of the employed oscillator frequency $\omega$ for
different sizes of the model space ($N+1$ denotes the number of
oscillator shells.). The convergence with respect to increasing size
of the model space is slow, and we are unable to achieve convergence
in up to 14 major oscillator shells. The results obtained in the
largest model spaces also show that the V$_{\mathrm{UCOM}}$ interaction
considerably overbinds $^{40}$Ca.  Within the CCSD approximation, we
find a ground-state energy of about $-133$~MeV for $^{16}$O and of about
$-400$ MeV for $^{40}$Ca.  The inclusion of triples corrections
within the $\Lambda$-CCSD(T) approximation yields a ground-state
energy of $-140.99$~MeV for $^{16}$O in 13 oscillator shells and an
oscillator frequency $\hbar\omega = 42$~MeV.

\begin{figure}[h]
\includegraphics[width=0.45\textwidth,clip=]{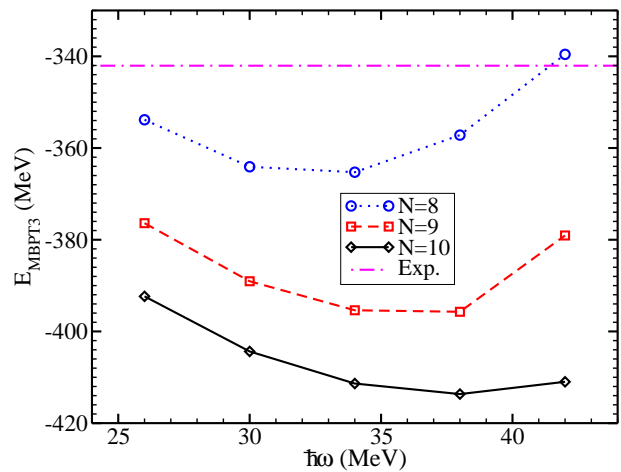}
\caption{(Color online) Ground-state energy for $^{40}$Ca using
  many-body perturbation theory to third order in the interaction
  $V_{{\rm UCOM}}$. The results are presented as a function of the
  oscillator spacing $\hbar\omega$ and the size of the model space
  $N$. As in the CCSD calculations, a Hartree-Fock basis has been
  used.}
  \label{fig:mbpt}
\end{figure}

We checked our results as follows. First, our CCSD results agree with
those reported for $^4$He in Ref.~\cite{Roth09} within about 50~keV in
model spaces of up to eight oscillator shells. Second, we also
computed the ground-state energy of $^{40}$Ca in third order many-body
perturbation theory.  The relevant diagrams which are included can be
found in Ref.~\cite{Dean04}.  Our results are shown in
Fig.~\ref{fig:mbpt} as function of the oscillator energy $\hbar\omega$
and the number of shells $N$. The results from many body perturbation
theory confirm the trend seen in the coupled-cluster calculations,
i.e., they exhibit a rather slow convergence in terms of the number of
shells $N$ and overbind the nucleus $^{40}$Ca. Reference~\cite{Roth06} also
solved the UCOM potential within many-body perturbation theory, albeit
for oscillator frequencies below $\hbar\omega\approx 20$~MeV or so. In
this regime, the binding energy is closer to the experimental value as
the small oscillator frequency acts as a cutoff in momentum space.

We speculate that the V$_{\mathrm{UCOM}}$ interaction still exhibits
relatively long tails in momentum space, which prevent a decoupling of
low- and high-momentum modes, or that peculiarities of the
corresponding cutoff yield a slow convergence of the UCOM interaction
in an oscillator basis. This speculation is borne out by two
observations. First, the comparison of the UCOM interaction with the
SRG interaction in Fig.~8 of Ref.~\cite{RothReinHer} shows that the
UCOM interaction does not decouple well in an oscillator basis. Second,
the energy minima in Figs.~\ref{fig:fig15}, \ref{fig:fig16} and
\ref{fig:mbpt} shift to higher frequencies $\hbar\omega$ as the size
of the model space is increased, as one would expect based on the
estimate~(\ref{hbaromega}) for an interaction with a large momentum
cutoff.  Our results also suggest that -- similar to other SRG
interactions and low-momentum interactions -- the UCOM interaction has to
be augmented by sizeable many-body forces~\cite{jpg2010}. The need for
these forces is not seen if one restricts the calculations to fixed
model spaces where model space parameters such as the oscillator
frequency and the number of shells introduce additional momentum
cutoffs.

\section{Summary}

We presented a spherical formulation of the coupled-cluster method for
the computation of energy spectra in nuclei with closed subshells and
their neighbors. This method has been used to solve nuclear structure
problems in model spaces consisting of up to 20 oscillator shells,
which allows us to obtain well-converged results for nucleon-nucleon
interactions derived from chiral effective field theory. We find that
chiral NN interactions saturate nuclei such as $^{16}$O and $^{48}$Ca
within 0.5~MeV per nucleon compared to data. The
two-particle--two-hole clusters provide about 90\% of correlation
energy, with the approximation of triples clusters accounting for the
remaining 10\%. We investigated the shell evolution in neutron-rich
isotopes of oxygen and fluorine and found that nucleon-nucleon
interactions alone fail to describe the experimentally observed
(sub)shell structure.

We also employed similarity renormalization group transformations of a
``bare'' chiral interaction. At the considered resolution scale, we
obtained overbinding of up to several MeV per nucleon. Again, the
missing contributions (when compared to experiment) are of natural
size in the power counting of effective field theory. These results
suggest that the systematic approach to nuclear structure can (in
principle) be extended to medium-mass nuclei. We studied the
$G$-matrix approach and found a very weak starting-energy dependence
in large model spaces combined with a rather slow convergence with
respect to increases in the size of the model space. For the UCOM
interaction, we find a slow convergence with respect to the size of the
model space and an overbinding for $^{40}$Ca that is comparable to
other low-momentum interactions.

We presented an in-depth study of the center-of-mass problem, and
demonstrated that the wave functions of the intrinsic Hamiltonian
factorize to a very good approximation into an intrinsic wave function
and a Gaussian for the center of mass. While an analytical
understanding of this behavior is still lacking, our calculations for
ground and excited states and for a virtually exactly solvable toy
model indicate that calculations in sufficiently large model spaces --
when based on the intrinsic Hamiltonian -- do not suffer from a
center-of-mass problem.

\section*{Appendix}

\section{Coupled-cluster  diagrams in a $J$-coupled scheme}
In this section we present expressions for several diagrams involved 
in our computation of the CCSD or $\Lambda$-CCSD(T) equations that 
involve various one-body, two-body, and three-body  amplitudes and operators. 
\subsection{Examples of CCSD diagrams in a $J$-coupled scheme}
We employ the following shorthands for the $T_1$ and $T_2$ amplitudes, 
namely  $t_{i}^{a}$ and $t_{ij}^{ab}$, respectively. To the latter we 
add the total two-particle angular momentum $J_{ij}$ in order 
to indicate a two-body wave operator with hole states $ij$ coupled to 
a two-body angular momentum $J_{ij}$, that is
\begin{eqnarray}
\nonumber
\lefteqn{t_{ij}^{ab}(J_{ij})=\langle (j_aj_b) J_{ij} | t | (j_ij_j)J_{ij} \rangle} \\
\nonumber
& & = \sum_{m_am_bm_im_j}C_{m_am_bM}^{j_aj_bJ_{ab}}C_{m_im_iM}^{j_ij_jJ_{ij}} \\
& & \langle (j_am_a)(j_bm_b)|t|(j_im_i)(j_jm_j)\rangle \ ,
\end{eqnarray}
with $m_i,m_j$, etc. being the magnetic quantum numbers  of the corresponding 
single-particle angular momenta $j_i,j_j$, etc. The coefficients $C$ are the 
standard Clebsch-Gordan coefficients.

The two-body amplitudes are diagonal in the total angular momentum, that is,
we have $J_{ij}=J_{ab}$.  The final matrix elements are independent of $M$.  
The labels $a,b,c,d\dots$ refer to particle states while $i,j,k,l,\dots$ 
are hole states.
  
\begin{figure}[h]
\includegraphics[width=0.30\textwidth,clip=]{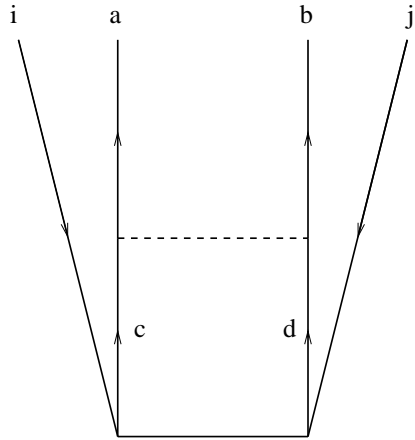}
\caption{(Color online) Diagram contributing to the $t_{ij}^{ab}(pp)$
  amplitudes in particle-particle coupled scheme.}
\label{fig2}
\end{figure}

In Fig.~\ref{fig2} we give a diagram contributing to the $T_2$
equation, and the corresponding algebraic expression in the $J-$coupled
scheme is (we list only the angular momentum related part), 
\[
\tilde{t}_{ij}^{ab}(J_{ij}) \leftarrow \sum_{cd}\langle (j_aj_b)J_{ab} | V | (j_cj_d)J_{ab} \rangle
\langle (j_cj_d) J_{ab} | t | (j_ij_j)J_{ab} \rangle \ .
\]
The above expression is obtained by identifying the diagram as a 
ladder cutting apart the lines connecting the various vertices. 
This technique is described in detail in Ref.~\cite{ellis1981}. 
Several applications of this technique are listed
also in Ref.~\cite{mhj95}. This is the most expensive diagram 
in the CCSD approximation and scales as $n_o^2n_u^4$, where $n_o$ 
is the number of occupied $j-$orbitals and $n_u$ is the number of 
unoccupied $j$-orbitals. As
seen, the diagram can be calculated by a matrix-matrix multiplication
utilizing efficient BLAS routines \cite{blas}.
\begin{figure}[h]
\includegraphics[width=0.30\textwidth,clip=]{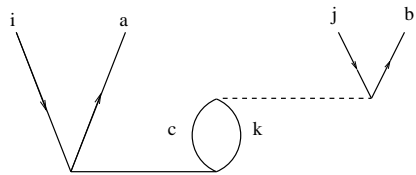}
\caption{(Color online) Diagram contributing to the $t_{ij}^{ab}(ph)$
  amplitudes in particle-hole coupled scheme.}
\label{fig3}
\end{figure}
The diagram in Fig.~\ref{fig3} also contributes to the $T_2$ equation,
and is easiest calculated in a particle-hole coupled scheme so that
intermediate sums of $9j-$symbols are avoided. The transformation of
the $t_{ij}^{ab}(J_{ij})$ amplitudes to a particle-hole coupled scheme
is given by
\begin{eqnarray}
\nonumber
\lefteqn{\langle (j_aj_i^{-1}) J_{ai} | t | (j_jj_b^{-1})J_{ai} \rangle = } \\
\nonumber
& & \sum_{J_{ab}}(2J_{ab} + 1)(-1)^{j_a+j_j+J_{ab}+J_{ai}}
\left\{\begin{array}{ccc} j_i & j_a & J_{ai} \\ j_b & j_j & J_{ab} \end{array}\right\}\times \\
& & \langle (j_aj_b) J_{ab} | t | (j_ij_j)J_{ab} \rangle \ ,
\end{eqnarray}
and the transformation back to particle-particle coupling is
\begin{eqnarray}
\nonumber
\lefteqn{\langle (j_aj_b) J_{ab} | t | (j_ij_j)J_{ab} \rangle = } \\
\nonumber
& & \sum_{J_{ai}}(2J_{ai} + 1) (-1)^{j_a+j_j+J_{ab}+J_{ai}}  
\left\{\begin{array}{ccc} j_i & j_j & J_{ab} \\ j_b & j_a & J_{ai} \end{array}\right\}\times \\
& & \langle (j_aj_i^{-1}) J_{ai} | t | (j_jj_b^{-1})J_{ai} \rangle \ .
\end{eqnarray}
The diagram in Fig.~\ref{fig3} can now easily be calculated in
a particle-hole coupled scheme as, 
\begin{eqnarray}
\nonumber
\lefteqn{\langle (j_aj_i^{-1}) J_{ai} | \tilde{t} | (j_jj_b^{-1})J_{ai} \rangle =} \\
\nonumber
& & \sum_{ck}(-1)^{j_k+j_c - J_{ai}+1} 
\langle (j_aj_i^{-1}) J_{ai} | t | (j_kj_c^{-1})J_{ai} \rangle \times \\
& & \langle (j_kj_c^{-1}) J_{ai} | V | (j_jj_b^{-1})J_{ai} \rangle \ ,
\end{eqnarray}
where again an efficient BLAS \cite{blas} matrix-matrix multiplication
routine can be used.

\begin{figure}[h]
\includegraphics[width=0.3\textwidth,clip=]{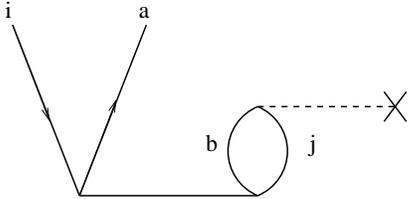}
\caption{(Color online) Diagram contributing to the $t_i^a$ amplitudes.}
\label{fig1}
\end{figure}

In Fig.~\ref{fig1} we give a diagram contributing to the $T_1$
equation.
The algebraic expression of Fig.~\ref{fig1} in a $J$-coupled scheme is,
\[ 
\tilde{t}^a_i \leftarrow \sum_{J_{ik}} {2J+1\over 2j_a + 1} \sum_{ck}
t_{ac}^{ik}(J_{ik}) \chi _k^b \ ,
\]
where $\chi _k^b$ is a one-body operator.

\subsection{Examples of $\Lambda$-CCSD(T) diagrams in a $J$-coupled scheme}

\begin{figure}[hbtp]
\includegraphics[width=0.45\textwidth,clip=]{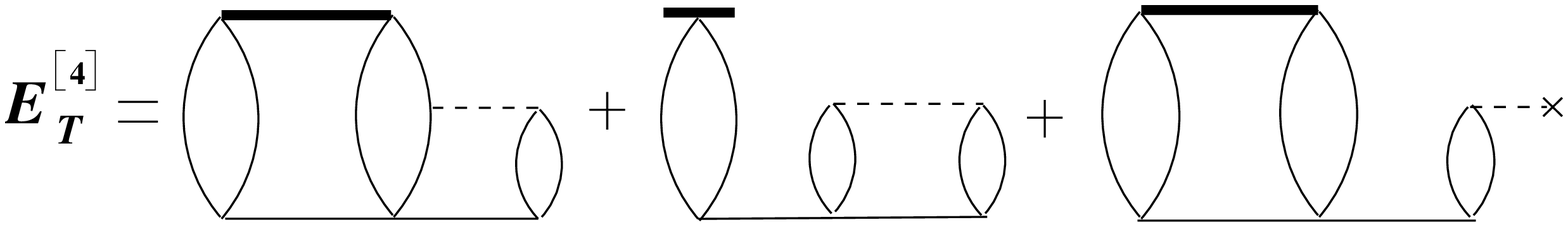}
\caption{(Color online) Diagrams for the $\Lambda$-CCSD(T) approximation.}
\label{fig:lambda1}
\end{figure}

\begin{figure}[hbtp]
\vspace{0.5cm}
\includegraphics[width=0.48\textwidth,clip=]{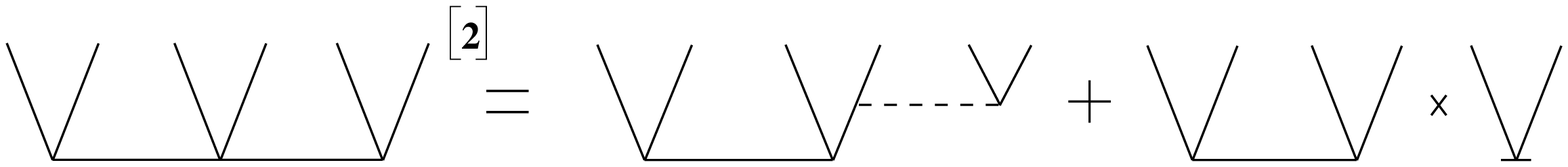}
\caption{(Color online) Diagrams for the $\Lambda$-CCSD(T) approximation.}
\label{fig:lambda2}
\end{figure}

Figure~\ref{fig:lambda1} shows an example of diagrams with a three-body
wave operator that give rise to contributions to the final energy.
These diagrams arise from the so-called $\Lambda$-CCSD(T) approximation
discussed in this work. 
The lower part of these diagrams is given by the three-body
$t$-amplitude labeled $t_{ijk}^{abc}$. It is antisymmetrized and in
the $\Lambda$-CCSD(T) approximation it is represented by a connected
three-body contribution consisting of a two-body amplitude
$t_{ij}^{ab}$ and a two-body interaction vertex plus a two-body
amplitude part $t_{ij}^{ab}$ times a one-body amplitude $t_{k}^{c}$.
These two contributions are depicted by the diagrams to the right of
the equality sign in Fig.~\ref{fig:lambda2}. The three-body amplitude
$t_{ijk}^{abc}$ is shown to the left of the equality sign in
Fig.~\ref{fig:lambda2}.

\begin{figure}[h]
\includegraphics[width=0.38\textwidth,clip=]{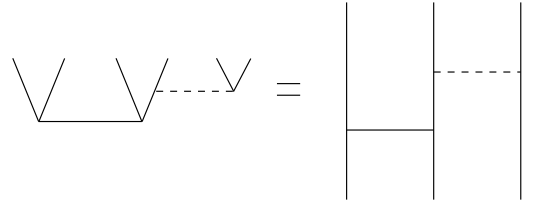}
\caption{(Color online) Connected part of three-particle-three-hole diagrams.}
\label{fig:lambda3}
\end{figure}

A diagram like the first to the right of the equality sign in 
Fig.~\ref{fig:lambda2} can be redrawn as shown in Fig.~\ref{fig:lambda3}.
Here we have stretched the hole lines in the diagram to the right of
the equality sign. Such a diagram with, for example, an intermediate
particle state can easily be calculated in an angular momentum coupled
scheme. The expressions can be derived using the methods discussed in
Ref.~\cite{ellis1981}. All three-body diagrams like those shown in
Fig.~\ref{fig:lambda3} have been listed in Ref.~\cite{polls1983}.

\begin{figure}[h]
\includegraphics[width=0.40\textwidth,clip=]{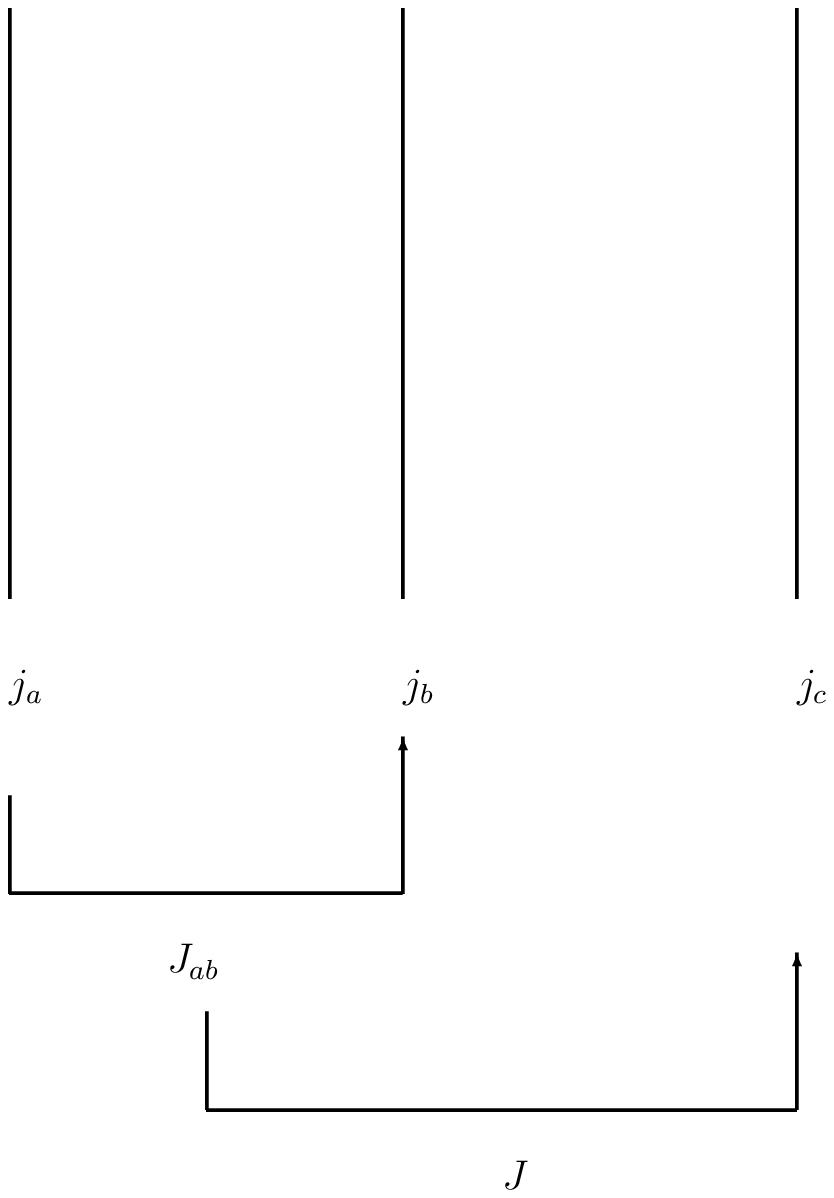}
\caption{The coupling order $([j_a\rightarrow j_b]J_{ab}\rightarrow j_c) J$. \label{fig:abc}}
\end{figure}

In order to derive these expressions, one needs to specify the given
coupling order for the angular momenta. Here we choose to couple our
angular momenta as
\begin{eqnarray}
\nonumber
\lefteqn{ | ([j_a\rightarrow j_b]J_{ab}\rightarrow j_c) J\rangle = } \\
\nonumber
& & \sum_{m_am_bm_c}\langle j_am_aj_bm_b|J_{ab}M_{ab}\rangle \langle J_{ab}M_{ab}j_cm_c|JM\rangle \\
& & |j_am_a\rangle\otimes |j_bm_b\rangle \otimes |j_cm_c\rangle \ , \label{eq:fabc}
\end{eqnarray}
and indicated in Fig.~\ref{fig:abc}.

Note that the two-body intermediate state is antisymmetric but
not normalized, that is, the state which involves the quantum numbers 
$j_a$ and $j_b$. We will hereafter assume that the intermediate 
two-body state is antisymmetric. With this coupling order, we can 
rewrite the general three-particle Slater determinant as 
\begin{equation}
\Phi(1,2,3) = {\cal A} | ([j_a\rightarrow j_b]J_{ab}\rightarrow j_c) J\rangle, 
\end{equation}
with an implicit sum over $J_{ab}$.  The final Slater determinant is
\begin{eqnarray}
\nonumber
\lefteqn{{\cal A} | ([j_a\rightarrow j_b]J_{ab}\rightarrow j_c) J\rangle = } \\
\nonumber 
& & \frac{1}{\sqrt{3!}}| ([j_a\rightarrow j_b]J_{ab}\rightarrow j_c) J\rangle- \\
\nonumber
& & \frac{1}{\sqrt{3!}}\left[ \sum_{J_{ac}}(-1)^{j_b+j_c+J_{ab}+J_{ac}}{\hat{J}_{ab}}{\hat{J}_{ac}} 
\left\{\begin{array}{ccc} j_a & j_b & J_{ab} \\ J & j_c & J_{ac} \end{array} \right\}+  \right. \\
\nonumber 
& & \left. \sum_{J_{bc}}(-1)^{J_{bc}}{\hat{J}_{ab}}{\hat{J}_{bc}}  \left\{\begin{array}{ccc} j_a & j_b & J_{ab} \\ 
J & j_c & J_{bc} \end{array} \right\} \right] | ([j_a\rightarrow j_b]J_{ab}\rightarrow j_c) J\rangle.
\end{eqnarray}
Here, we used the shorthand $\hat{J}\equiv \sqrt{2J+1}$. 
With this coupling order, one can then compute all three-body diagrams
like those listed in Fig.~\ref{fig:lambda3}.

\begin{figure}[h]
\includegraphics[width=0.40\textwidth,clip=]{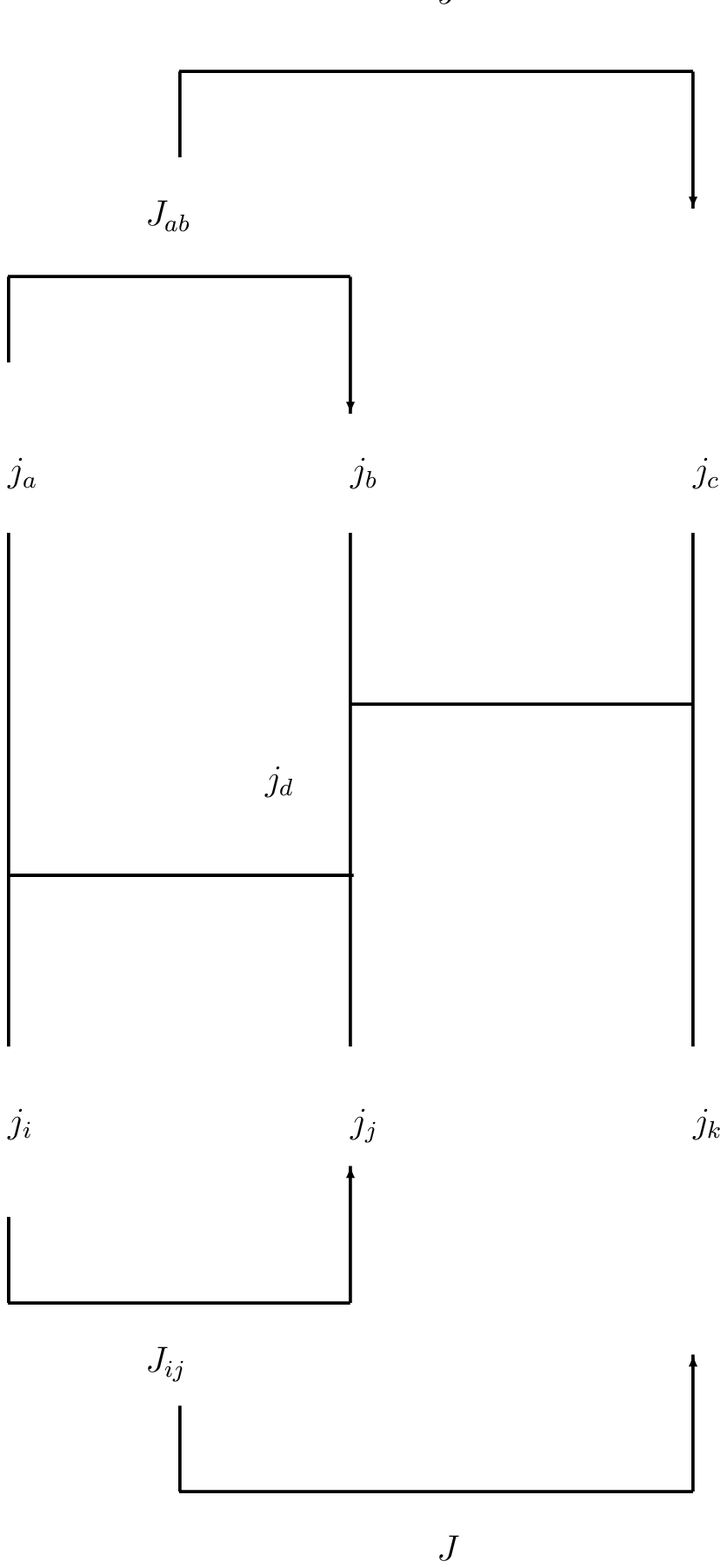}
\caption{Three-body diagram with two-body vertex and amplitude and a particle 
state as intermediate single-particle states. The sum over $d$ runs over all 
possible quantum numbers of the intermediate particle(hole) state. The algebraic expression 
of the above diagram, in an angular momentum coupled scheme, is given in Eq.~(\ref{eq:eqthreebody}).
\label{fig:threebodyparticle}}
\end{figure}

Here we give an example of the closed form expression for the angular
momentum recoupled part of a selected three-body diagram with two-body
amplitudes $t_{ij}^{ab}$. These are diagrams which start with a
three-body state but have two-body wave operators and end in a
contribution to a three-body state with a two-body vertex after the
two-body amplitude, as shown in Fig.~\ref{fig:threebodyparticle}.
Again, we employ the shorthand $t_{ij}^{ab}(J_{ij})$ to indicate a
two-body wave operator with hole states $ij$ coupled to a two-body
angular momentum $J_{ij}$, that is, $t_{ab}^{ij}(J_{ij})=\langle
(j_aj_b) J_{ij} | t | (j_ij_j)J_{ij} \rangle$.  The diagram with a
particle intermediate state is shown in
Fig.~\ref{fig:threebodyparticle} with its angular momentum
representation,
\begin{eqnarray}
\nonumber 
\lefteqn{\sum_{J_{bc}}\sum_d(-1)^{j_b+j_c+j_d+J_{bc}}{\hat{J}_{ab}}{\hat{J}_{ij}}{\hat{J}_{bc}}^2
\left\{\begin{array}{ccc} j_a & j_b & J_{ab} \\ j_c & J & J_{bc} \end{array}\right\}} \\
& & \left\{\begin{array}{ccc} j_a & j_d & J_{ij} \\ j_k & J & J_{bc} \end{array}\right\} 
\langle (j_bj_c) J_{bc} | V | (j_dj_k)J_{bc} \rangle t_{ad}^{ij}(J_{ij}).
\label{eq:eqthreebody}
\end{eqnarray}
The expression for this diagram is obtained by opening up the
intermediate three-particle state $j_aj_dj_k$ and recoupling the
angular momenta $j_b$ and $j_c$ to yield a final two-particle angular
momentum $J_{bc}$.  This applies also to the single-particle angular
momenta $j_d$ and $j_k$. These two single-particle angular momenta
couple to the same final two-particle angular momentum $J_{bc}$. These
recouplings are reflected in the two $6j$-symbols in the above
expression.  The expression for the corresponding diagram of
Fig.~\ref{fig:threebodyparticle} with a hole intermediate state is
obtained by replacing the particle labeling $j_d$ with $j_l$. The
subscript $l$ refers to the fact that this is a single-hole state. The
internal hole line, however, gives rise to a factor $-1$.

In total there are, due to the antisymmetry of the three-body wave
function, nine diagrams with an intermediate particle state and nine
diagrams with an intermediate single-hole state.  The angular momentum
expressions for these diagrams are listed in Ref.~\cite{polls1983}.
The expressions for diagrams like the rightmost one in
Fig.~\ref{fig:lambda2} yield similar expressions and are easy to
compute. They represent a two-body part coupled with a one-body part.

We thank S.K.~Bogner, R.J.~Furnstahl, and A. Schwenk for useful
discussions.  This work was supported by the Office of Nuclear Physics, 
U.S. Department of Energy (Oak Ridge National Laboratory); the University of Washington 
under Contract No. DE-FC02-07ER41457; and the University of Tennessee 
under Grant No.\ DE-FG02-96ER40963. 
MHJ thanks the Research Council of Norway for support.

\end{document}